\documentclass[aps,twocolumn,pra,superscriptaddress,amsmath,amssymb]{revtex4-2} %% for PRL
\usepackage{dcolumn}
\usepackage{bm}
\usepackage{amsmath}
\usepackage{txfonts}
\usepackage[T1]{fontenc}
\usepackage{xspace}
\usepackage{ulem}
\usepackage{comment}
\newcommand{\bras}[1]{\langle#1\rvert}
\newcommand{\kets}[1]{\lvert#1\rangle}

\newcommand{\means}[1]{\langle#1\rangle}

\newcommand{\meanss}[1]{\langle\!\langle#1\rangle\!\rangle}

\ifx\pdfoutput\undefined
\usepackage[dvipdfmx]{graphicx}
\usepackage[dvipdfmx]{hyperref}
\usepackage[dvipdfmx]{color}
\usepackage[dvipdfmx]{xcolor}
\else
\usepackage{graphicx}
\usepackage{hyperref}
\usepackage{color}
\usepackage{xcolor}
\fi
\hypersetup{colorlinks=true,linkcolor=blue,citecolor=blue,urlcolor=blue}
\begin{document}
\let\emph\textit

\title{
  Real-time control of non-Abelian anyons in Kitaev spin liquid under energy dissipation
}
\author{Chihiro Harada}
\affiliation{
  Department of Physics, Tohoku University, Sendai 980-8578, Japan
}
\author{Atsushi Ono}
\affiliation{
  Department of Physics, Tohoku University, Sendai 980-8578, Japan
}
\author{Joji Nasu}
\affiliation{
  Department of Physics, Tohoku University, Sendai 980-8578, Japan
}

\date{\today}
\begin{abstract}
Quantum spin liquids realized in the Kitaev model offer a platform for fractionalization of spin into two quasiparticles: itinerant Majoranas and localized visons.
Introducing a uniform weak magnetic field associates a Majorana zero mode with each vison excitation.
The vison accompanied by a Majorana zero mode is known to behave as a non-Abelian anyon, which has garnered significant attention for its potential applications in topological quantum computing.
Although spatial and temporal control of these anyons is essential for exploring their applicability in quantum computing, numerical simulations of creating, moving, and annihilating non-Abelian anyons by an external field remain challenging as this field violates the exact solvability of the Kitaev model.
Moreover, such a field to control non-Abelian anyons may disturb the quantum state due to the energy injection it causes.
In this study, by introducing energy dissipation phenomenologically in real-time simulations, we demonstrate that the generation, movement, and annihilation of vison excitations can be achieved while maintaining their localization.
In particular, we find that a vison can be moved in a desired direction by using time-dependent local magnetic fields or gradient magnetic fields, and it remains accompanied by a Majorana zero mode even after its movement.
We also reveal that a larger spatial extent of the Majorana zero modes bound to a vison facilitates the movement of the vison with smaller field gradients.
Furthermore, our numerical simulations demonstrate pair creation and annihilation of visons, triggered by time-dependent magnetic fields.
The results obtained in this study highlight the significance of energy dissipation in controlling non-Abelian anyons, which will stimulate further investigations into the nonequilibrium dynamics of fractional quasiparticles in strongly correlated electron systems, as well as studies for applications in topological quantum computation.
\end{abstract}
\maketitle

%%%%%%%%%%%
% Introduction

\section{Introduction}

In condensed matter physics dealing with strong electron correlations, exotic quantum many-body ground states potentially provide a playground for emergent quasiparticle excitations that are entirely different from electrons. 
Among them, topologically nontrivial nature inherent to ground states are known to give rise to exotic quasiparticles such as Majorana fermions and non-Abelian anyons. 
Majorana fermions, particles that are identical to their own antiparticles~\cite{Majorana1937}, have garnered interest for their zero-energy states in materials, known as Majorana zero modes, and have been studied primarily in topological superconductors~\cite{Sarma2006,Sato2009topological,Fu2008Superconducting,Sato2016Majorana,Sato_rev2017} and quantum nanowires~\cite{Lutchyn2010,Oreg2010,mourik2012signatures,das2012zero,albrecht2016exponential}.
These modes can be localized at vortices or edges and are anticipated to behave as non-Abelian anyons~\cite{Wilczek1982,moore1991nonabelions}.
Non-Abelian anyons, also discussed in the context of fractional quantum Hall systems, are considered to have unique statistical properties originating from quantum many-body effects~\cite{moore1991nonabelions,Stormer1999,Read2000,banerjee2018observation}.
Since it is possible to achieve topological quantum computation by braiding multiple anyons spatially, non-Abelian anyons have attracted significant attention not only in condensed matter physics but also in the field of quantum information~\cite{kitaev2003fault,freedman2003topological,Nayak2008,Ahlbrecht2009quantum_computer}.

The Kitaev quantum spin liquid (QSL) is another quantum many-body state exhibiting non-Abelian anyonic quasiparticles.
This state is realized as the ground state of the Kitaev spin model, which is an exactly solvable quantum spin model describing interactions between $S=1/2$ spins on a honeycomb lattice~\cite{Kitaev2006}.
In the Kitaev QSL, elementary excitations of a quantum spin are fractionalized into two quasiparticles: Majorana fermions and visons, the latter being vortex-like local $Z_2$ excitations~\cite{RevModPhys.87.1,Hermanns2018rev,Knolle2019rev,takagi2019rev,Janssen_2019rev,Motome2020rev,Trebst2022rev}.
The Majorana fermions are itinerant on the honeycomb lattice and exhibit gapless linear dispersions.
Introducing weak magnetic fields opens a gap and induces a topologically nontrivial Majorana state.
In this situation, a Majorana zero mode appears around each vison excitation.
A composite excitation coupled between a vison and the Majorana zero mode localized at the vison behaves as a non-Abelian anyon~\cite{Kitaev2006}.

The Kitaev QSL holds inherent potential for applications in topological quantum computing using non-Abelian anyons, which motivates substantial efforts to realize the Kitaev model in various materials.
Among these efforts, strongly correlated electron systems with strong spin-orbit interactions have been proposed as promising candidates~\cite{PhysRevLett.102.017205}, in particular, iridium oxides~\cite{PhysRevB.82.064412,PhysRevLett.108.127203,PhysRevLett.108.127204,PhysRevLett.109.266406,Mehlawat2017heat-capacity} and the ruthenium compound $\alpha$-RuCl$_3$~\cite{PhysRevB.90.041112,PhysRevB.91.094422,PhysRevLett.114.147201,Johnson2015,PhysRevB.91.144420,Cao20016,Nasu2016nphys,Do2017majorana,janvsa2018observation,banerjee2018excitations,Balz2019,Widmann2019,wang2020range}.
For the emergence of non-Abelian anyons, the itinerant Majorana fermion system needs to be topologically nontrivial, which is theoretically predicted to be verified through the half-integer quantization of a thermal Hall conductivity~\cite{Kitaev2006,Nasu2017}.
Recently, there has been considerable progress in measuring a thermal Hall effect in $\alpha$-RuCl$_3$~\cite{Kasahara2018}, with discussions focusing on the half-integer quantization~\cite{kasahara2018majorana,Yamashita2020_Sample_dependence,bruin2022robustness}, the field-angle dependence of the thermal Hall conductivity and specific heat~\cite{Koyama2021,Zhang2021,hwang2022identification,tanaka2022thermodynamic,Imamura2024}, and contributions from other quasiparticles~\cite{Hentrich2019,czajka2022planar,Lefrancois2022}.
Meanwhile, in iridium and ruthenium compounds, the presence of Heisenberg and $\Gamma$ interactions, in addition to the Kitaev interaction, has been identified~\cite{PhysRevLett.105.027204,PhysRevLett.110.097204,PhysRevLett.113.107201,PhysRevLett.112.077204,Winter2016,Song2016,winter2017breakdown,Gohlke2018dynamics}.
Suppressing additional interactions other than the Kitaev interaction is considered crucial for realizing non-Abelian anyons.
To reduce these additional interactions, substantial efforts have been made to explore new Kitaev platforms.
% Furthermore, the exploration of new materials is being actively pursued.
The theoretical possibility of realizing the Kitaev model has also been proposed in other transition metal compounds~\cite{bette2017solution3LiIr2O6,Kitagawa2018nature,Geirhos2020H3LiIr2O6,Haraguchi2018,miura2020stabilization,Jang2021ilmenites,Nawa2021RuX3,kim2021RuX3,Imai2022RuX3,Liu2018Pseudospin,sano2018,Jang_AFKitaev2019}, their thin films~\cite{Zhou2019,mashhadi2019spin,Kohsaka2024pre}, organic compounds~\cite{Yamada2017}, cold atom systems~\cite{Duan2003,Manmana2013,gorshkov2013kitaev,Fukui2022}, and superconducting circuits~\cite{Sameti2019}.

In the Kitaev model, the implementation of topological quantum computation requires the observation, control, and generation of vison excitations, which can behave as non-Abelian anyons when each of them is accompanied by a Majorana zero mode.
Recently, theoretical proposals have been made to observe vison excitations and Majorana zero modes using scanning tunneling microscopy (STM)~\cite{Feldmeier2020,Konig2020,Pereira2020,Chen2020Impurity,Carrega2020,Udagawa2021,Bauer2023,Takahashi2023} and interferometry~\cite{Klocke2021,Wei2021,Liu2022AnyonGeneration}.
Motivated by these proposals, experimental imaging of thin films of Kitaev candidate materials using STM has also been conducted~\cite{Kohsaka2024pre}.
Furthermore, vison control using atomic force microscopy (AFM)~\cite{Jang2021} and local magnetic fields~\cite{Harada2023} have been proposed in Kitaev QSLs.
Additionally, in the Kitaev ladder system, it has been suggested that a quantum state capable of possessing Majorana zero modes can be selected by projective measurements~\cite{Xu2024pre}.
Since the quantum state changes during these processes in the time domain, any attempts to move the visons inject energy into the system.
This can delocalize vison excitations, preventing their application in topological quantum computation.
Therefore, it is highly desirable to propose protocols for controlling visons while maintaining their localization and the presence of Majorana zero modes associated with them.

In this paper, we demonstrate the movement, creation, and annihilation of non-Abelian anyons in a Kitaev quantum spin liquid under energy dissipation, which is introduced to enhance the localization of vison excitations and improve their controllability.
We perform real-time simulations driven by temporally and spatially dependent external fields, starting with a state where visons are excited in a quantum spin system described by the Kitaev model, using time-dependent Hartree-Fock theory with spatially dependent mean fields.
In this theory, the time dependence of the density matrix obeys the von Neumann equation, with energy dissipation introduced by adding a term relaxing the system toward the ground state of an instantaneous mean-field Hamiltonian. 
Numerical simulations for vison manipulations by applying pulsed external fields show that introducing energy dissipation significantly suppresses the spread of visons.
We find that a Majorana zero mode moves along with a vison.
Furthermore, we reveal that the movement of visons is induced by the spatial modulation of the external field, sensed by the spatially extended Majorana zero modes.
Such vison driving can occur not only by locally applied pulsed magnetic fields near an excited vison but also by gradient magnetic fields.
Particularly, we clarify that the driving of visons becomes easier when the topological gap of the Majorana system is small and the spread of the Majorana zero modes is large.
We also demonstrate that the pair creation and pair annihilation of visons occur by introducing a time-dependent magnetic field locally applied in the system.

This paper is organized as follows.
In the next section, we present the Kitaev model and its fundamental properties.
Section~\ref{sec:method} describes the method for the real-time simulations used in this study and the definitions of physical quantities.
The time-dependent mean-field theory is introduced in Sec.~\ref{sec:mf-theory}.
To incorporate energy dissipation, we adopt a relaxation-time approximation given in Sec.~\ref{sec:RTA}.
In Sec.~\ref{sec:quasiparticles}, we introduce the spatial distributions of vison excitations and the low-energy Majorana density of states (DOS) as time-dependent physical quantities.
Section~\ref{sec:result} presents the results obtained from real-time simulations of driving, annihilating, and creating visons.
Effects of energy dissipation on vison movement are clarified in Sec.~\ref{energy-dissipation}.
In Sec.~\ref{sec:size-dep}, we examine the size dependence of areas where a time-dependent magnetic field is applied for vison control.
The numerical results on vison driving by a gradient magnetic field are presented in Sec.~\ref{sec:gradient-field}.
Sections~\ref{sec:annihilation} and \ref{sec:pair-creation} present results on real-time simulations for pair annihilation and pair creation, respectively.
In Sec.~\ref{sec:discussion}, we discuss the relevance to experimental results and perspectives on this study.
Finally, Sec.~\ref{sec:summary} is devoted to the summary.

%%%%%%%%%%%
% Model
\section{Model}
\label{sec:model}

The Kitaev model under a uniform magnetic field is effectively written as~\cite{Kitaev2006}
\begin{align}
    {\cal H}_K=-J\sum_{\means{jj'}_\gamma}S_j^\gamma S_{j'}^\gamma -\kappa \sum_{\meanss{jj'j''}_{\gamma\gamma'\gamma''}}S_{j}^{\gamma} S_{j'}^{\gamma'} S_{j''}^{\gamma''},
    \label{eq:Kitaev-Hamil}
\end{align}
where $\bm{S}_j=(S_j^x,S_j^y,S_j^z)$ denotes an $S=1/2$ spin located at site $j$ on a honeycomb lattice, and the first term represents the ferromagnetic Kitaev-type interaction with $J$ between spins at $j$ and $j'$ sites on the $\gamma$ bond $\means{jj'}_{\gamma}$ with $\gamma=x,y,z$.
The second term represents an effective magnetic field with $\kappa$ obtained from the third-order perturbation expansion with respect to the Zeeman term arising from a uniform magnetic field.
This term is written by the product of three spins at neighboring sites $\meanss{jj'j''}_{\gamma\gamma'\gamma''}$ composed of two nearest neighbor bonds $\means{jj'}_{\gamma}$ and $\means{j'j''}_{\gamma''}$ with $\gamma'\ne \gamma,\gamma''$.
In this model, there is a conserved quantity $W_p$ on each hexagon plaquette $p$, which is defined by
\begin{align}
    W_p=2^6 \prod_{j\in p}S_j^\gamma.
\end{align}
Since $W_p^2=1$, all eigenstates of ${\cal H}_K$ is characterized by a configuration of $W_p=\pm1$, which we call a vison sector.
The ground state of the Kitaev model given by Eq.~\eqref{eq:Kitaev-Hamil} belongs to a subspace with $W_p=+1$ for all hexagon plaquettes in the honeycomb lattice.
The local excitation to $W_p=-1$ on a certain plaquette is regarded as a quasiparticle, called a vison.
The vison is a localized excitation, which does not move to other plaquettes within ${\cal H}_K$.

\begin{figure*}[t]
  \begin{center}
  \includegraphics[width=2\columnwidth,clip]{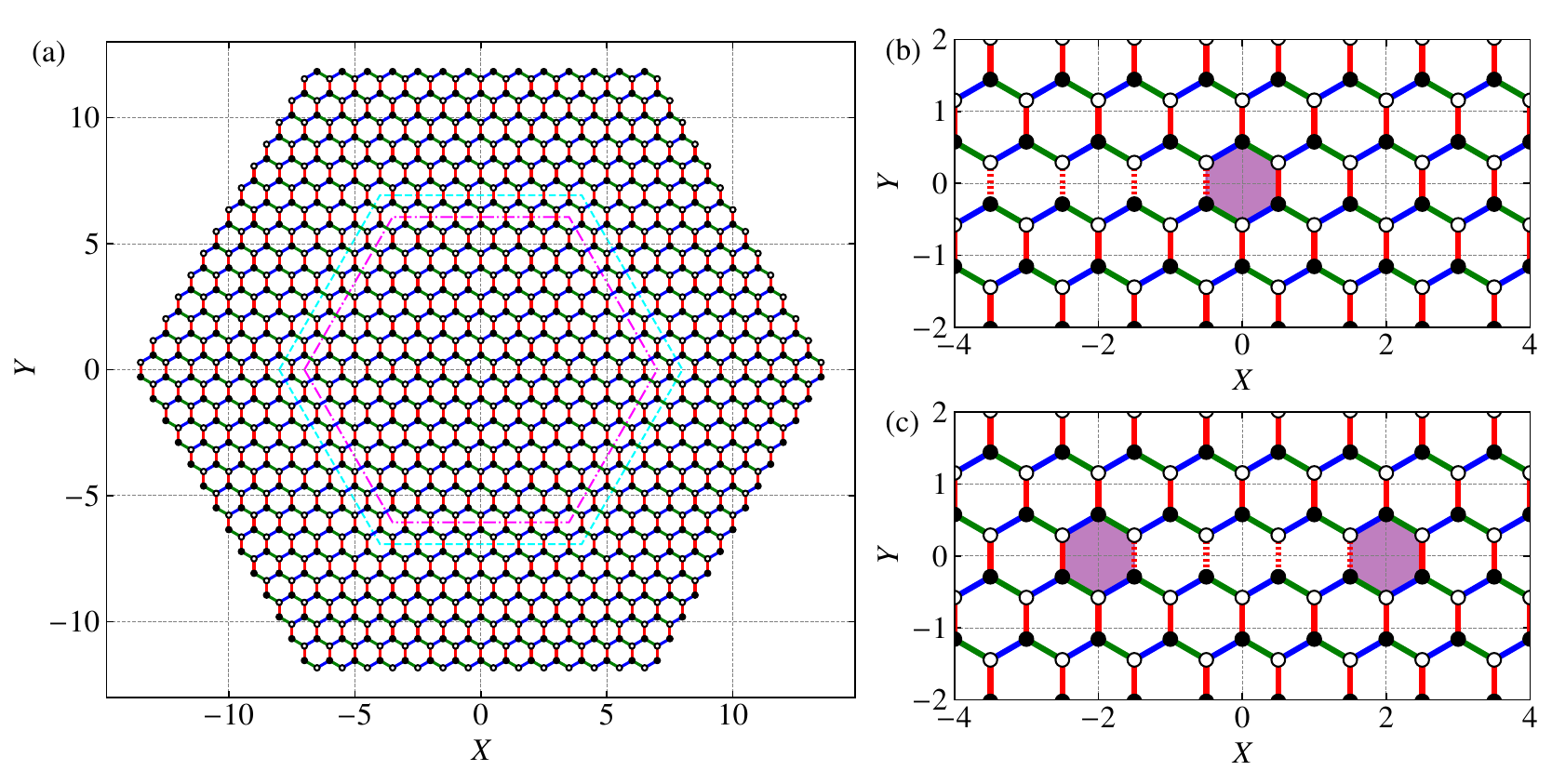}
  \caption{
  (a) Honeycomb lattice cluster with $N=1176$ sites used in the present calculations except for simulations on pair vison creations shown in Sec.~\ref{sec:pair-creation}.
  We perform simulations for the pair vison creations in the $N=294$ cluster, which is surrounded by the magenta dashed-dotted line.
  The cyan dashed line represents the area where a gradient magnetic field is applied in Sec.~\ref{sec:gradient-field}.
  Blue, green, and red lines represent the $x$, $y$, and $z$ bonds, respectively, and
filled (open) circles denote sites belonging to the A (B) sublattice.
(b), (c) Schematic figures of states with (b) a vison and (c) two visons, where $\eta_r$ on the $z$ bonds, denoted by dashed lines are $-1$.
  }
  \label{fig:lattice}
  \end{center}
\end{figure*}

In a fixed vison sector, the Kitaev model under the effective magnetic field can be written as a free Majorana fermion model by introducing two Majorana fermions $c_j$ and $\bar{c}_j$ at each site via the Jordan-Wigner transformation~\cite{PhysRevB.76.193101,PhysRevLett.98.087204,1751-8121-41-7-075001,PhysRevLett.113.197205,PhysRevB.92.115122}:
\begin{align}
  &{\cal H}_{K}=-\frac{J}{4}\sum_{[jj']_x}ic_j c_{j'}-\frac{J}{4}\sum_{[jj']_y}ic_j c_{j'}
  -\frac{J}{4}\sum_{[jj']_z}\eta_r ic_j c_{j'}\nonumber\\
  &-\frac{\kappa}{8}\sum_{\meanss{jj'j''}_{zxy}}\eta_r ic_j c_{j''}
  -\frac{\kappa}{8}\sum_{\meanss{jj'j''}_{xyz}}\eta_r ic_j c_{j''}
  -\frac{\kappa}{8}\sum_{\meanss{jj'j''}_{yzx}}ic_j c_{j''},
  \label{eq:Hamilc}
\end{align}
where $[jj']_\gamma$ is an ordered pair on a $\gamma$ bond and $\eta_r =i\bar{c}_j \bar{c}_{j'}$ on a $z$ bond $r$ with $j$ ($j'$) belonging to sublattice $A$ ($B$) [see Fig.~\ref{fig:lattice}(a)].
This term can also arise from $\Gamma'$ interactions~\cite{Takikawa2019,Takikawa2020} in addition to a uniform magnetic field~\cite{Kitaev2006}.
Note that $\eta_r$ is a local conserved quantity taking $\pm 1$, which is related to $W_p$ as
\begin{align}
    W_p=\prod_{r\in p} \eta_r.
\end{align}
From this relation, the state with $W_p=+1$ for all hexagon plaquettes is expressed by the state with $\eta_r=+1$ for all $z$ bonds.
Furthermore, an excited state with a vison is represented by flipping $\eta_r$ to $-1$ on the $z$ bonds intersecting with a half-line originating from the vison [see Fig.~\ref{fig:lattice}(b)], and the state with two visons is represented by $\eta_r=-1$ on the $z$ bonds intersecting with a line connecting the visons [see Fig.~\ref{fig:lattice}(c)].
In the case with nonzero $\kappa$, a zero-energy state consisting of the Majorana fermion $c$ emerges in the vicinity of an isolated vison.
This is referred to as a Majorana zero mode, which is distributed with a spread centered around the vison, and their radius increases with decreasing $\kappa$.

In the system described by Eq.~\eqref{eq:Kitaev-Hamil}, an excited vison remains stationary and does not change over time as $W_p$ is a conserved quantity even with nonzero $\kappa$.
To control a vison, one needs to introduce an external field that does not commute with $W_p$.
Here, we consider a time-dependent local magnetic field, which is given by
\begin{align}\label{eq:tdep-Hmag}
    {\cal H}_h(t)=-\sum_{j\in {\cal S}} h_j(t) S_j^z,
\end{align}
where $h_j(t)$ is an external field depending on both time and space applied in area ${\cal S}$.
The local field does not commute with $W_p$ with $p\in {\cal S}$, capable of manipulating a vison temporally and spatially.
Note that since $S_i^z$ flips two $W_p$ of two adjacent plaquettes separated by the $z$ bond connecting site $i$, the magnetic field in Eq.~\eqref{eq:tdep-Hmag} can move a vison along the $X$ direction.

%%%%%%%%%%%
% Method
\section{Method}
\label{sec:method}

\subsection{Majorana mean-field theory}
\label{sec:mf-theory}

To solve the time-dependent Hamiltonian ${\cal H}(t)={\cal H}_K+{\cal H}_h(t)$, we adopt Majorana mean-field theory.
We introduce Majorana fermions $\{\gamma_l\} = \{c_1, c_2, \cdots, c_N, \bar{c}_1, \cdots, \bar{c}_N\}$ with $l=1,2,\cdots,2N$ including the two Majorana fermions $c_j$ and $\bar{c}_j$, where $N$ is the number of sites.
When $h_j(t)$ is nonzero, $\eta_r$ is no longer a conserved quantity, and we apply the following decoupling to the third, fourth, and fifth terms of Eq.~\eqref{eq:Hamilc}~\cite{nasu2017ising,Nasu2018mag}:
\begin{align}\label{eq:MFdecoupling}
    \gamma_l \gamma_{l'} \gamma_{l''} \gamma_{l'''}\simeq&
    \means{\gamma_l \gamma_{l'}} \gamma_{l''} \gamma_{l'''}
    +\gamma_l \gamma_{l'} \means{\gamma_{l''} \gamma_{l'''}}
    -\means{\gamma_l \gamma_{l'}} \means{\gamma_{l''} \gamma_{l'''}}\notag\\
    &-
    \means{\gamma_l \gamma_{l''}} \gamma_{l'} \gamma_{l'''}
    -\gamma_l \gamma_{l''} \means{\gamma_{l'} \gamma_{l'''}}
    +\means{\gamma_l \gamma_{l''}} \means{\gamma_{l'} \gamma_{l'''}}\notag\\
    &+
    \means{\gamma_l \gamma_{l'''}} \gamma_{l'} \gamma_{l''}
    +\gamma_l \gamma_{l'''} \means{\gamma_{l'} \gamma_{l''}}
    -\means{\gamma_l \gamma_{l'''}} \means{\gamma_{l'} \gamma_{l''}}.
\end{align}
Using the above approximation, we obtain the mean-field Hamiltonian $\mathcal{H}_{\rm MF}(t)$ from ${\cal H}(t)$ as the following bilinear form of the Majorana fermions $\{\gamma_l\}$:
\begin{align}
\mathcal{H}_{\rm MF}(t) = \frac{i}{4}\sum_{ll'}\gamma_l\mathcal{A}_{ll'}\gamma_{l'} + C,
\label{eq:MF1}
\end{align}
where $C$ stands for terms not including $\{\gamma_l\}$, and $\mathcal{A}$ depends on time $t$ and mean fields $\means{\gamma_l \gamma_{l'}}$.

The time evolution is governed by the mean-field Hamiltonian $\mathcal{H}_{\rm MF}$.
We introduce the $2N\times 2N$ density matrix $\rho$ as a set of the mean fields $\means{\gamma_l \gamma_{l'}}=\bras{\Psi(t)}\gamma_l \gamma_{l'}\kets{\Psi(t)}$, which is given by 
\begin{align}
    [\rho(t)]_{ll'} =\frac{1}{2}\bras{\Psi(t)}\gamma_{l'} \gamma_{l}\kets{\Psi(t)},
\end{align}
where $\kets{\Psi(t)}$ is a many-body wave function at time $t$. 
This wave function is determined by the following Schr\"{o}dinger equation:
\begin{align}
    i\frac{\partial}{\partial t}\kets{\Psi(t)}=\mathcal{H}_{\rm MF}(t)\kets{\Psi(t)},
\end{align}
where we assume $\hbar=1$.
From this equation, the von Neumann equation that the density matrix obeys is given by~\cite{volkov1973collisionless,Tsuji_theoryof2015,Murakami_collective2020,Nasu2019_realtime,Minakawa2020,Nasu2022}
\begin{align}\label{eq:vNeq}
    \frac{\partial \rho}{\partial t}=[i\mathcal{A},\rho].
\end{align}
Since the matrix $\mathcal{A}$ is a function of the mean fields $\rho_{ll'}$, we solve this equation by the fourth-order Runge-Kutta method with time step $\Delta t$.
The mean-field Hamiltonian $\mathcal{H}_{\rm MF}(t)$ can be written as
\begin{align}
\mathcal{H}_{\rm MF}(t) = \sum_{\lambda=1}^{N}\varepsilon_\lambda\left(f_\lambda^\dagger f_\lambda-\frac12\right) + C,
\label{eq:MF2}
\end{align}
where $i\mathcal{A}$ is diagonalized with a $2N\times2N$ unitary matrix as
\begin{align}
  U^\dagger i{\cal A}U={\rm diag}\{\varepsilon_1,\varepsilon_2,\cdots, \varepsilon_N,-\varepsilon_1,-\varepsilon_2,\cdots, -\varepsilon_N\}.
\end{align}
Here, $\varepsilon_\lambda=-\varepsilon_{\lambda+N}\ge 0$, and $f_\lambda^\dagger$ ($f_\lambda$) is a creation (annihilation) operator of a fermion with energy $\varepsilon_\lambda$, which satisfies
\begin{align}
\gamma_l =\sqrt{2}\sum_{\lambda=1}^{2N} U_{l\lambda}f_{\lambda}
=\sqrt{2}\sum_{\lambda=1}^{N}
\left(U_{l\lambda}f_{\lambda} +  U_{l\lambda}^{*}f^{\dagger}_{\lambda}\right).
\end{align}

At the initial time $t=t_{\rm in}$ of time-dependent calculations, we assume $h_j(t_{\rm in})=0$.
In this case, $W_p$ and $\eta_r$ are local conserved quantities, and the mean-field decouplings in Eq.~\eqref{eq:MFdecoupling} hold exactly.
As an initial condition, we choose $\kets{\Psi(t_{\rm in})}=\kets{\Phi(t_{\rm in})}$, where $\kets{\Phi(t)}$ is the lowest energy state of $\mathcal{H}_{\rm MF}(t)$.
Note that the unitary matrix must fulfil $U_{l,\lambda+N}=U_{l\lambda}^{*}$ and the fermionic operators satisfy $f_{\lambda+N}=f_{\lambda}^\dagger$.
To enforce this condition even in the case with zero energy eigenvalues, we adopt the Schur decomposition $Q^T{\cal A}Q$ for a real skew-symmetric matrix ${\cal A}$ with a real orthogonal matrix $Q$ in numerical calculations~\cite{nasu2023rev}.

In real-time simulations, the initial time is chosen to be $t_{\rm in}=0$.
At this time, $\mathcal{H}_{\rm MF}(t_{\rm in})$ is identical to $\mathcal{H}_K$.
% We determine the initial configuration of visons by choosing $\{\eta_r\}$ appropriately.
The configuration of $\{\eta_r\}$ in $\mathcal{H}_K$ is appropriately selected according to each problem addressed in Sec.~\ref{sec:result}.
Then, we obtain the matrix elements of $\rho(t_{\rm in})$.
Based on this initial density matrix, we numerically solve the von Neumann equation in Eq.~\eqref{eq:vNeq}.
We set $\Delta t/J^{-1}=0.1$ and perform simulations on hexagon-shaped clusters with $N=1176$ shown in Fig.~\ref{fig:lattice}(a), except for numerical simulations on pair creation of visons in Sec.~\ref{sec:pair-creation}.

\subsection{Relaxation time approximation}
\label{sec:RTA}

In real systems, there exists energy dissipation due to couplings between the system and environment.
To incorporate this effect, we add a new term to Eq.~\eqref{eq:vNeq} as,
\begin{align}\label{eq:vNeq-diss}
    \frac{\partial \rho}{\partial t}=[i\mathcal{A},\rho] + I(\rho),
\end{align}
where $I(\rho)$ represents a coupling with environment, such as lattice degrees of freedom.
Since this term is introduced as energy dissipation, this term can reduce the energy of the system.
In the present calculations, we assume that the timescale for the system to relax to the ground state of an instantaneous mean-field Hamiltonian is shorter than the timescale for it to relax to the true ground state without visons due to energy dissipation.
Under this assumption, we only consider the former relaxation process and adopt the following form for $I(\rho)$ as a relaxation time approximation:
\begin{align}\label{eq:RTA}
    I^R(\rho)=-\frac{\rho(t)-\rho_g(t)}{\tau}
\end{align}
where $\tau$ is the relaxation time, and $\rho_g(t)$ the density matrix representing the ground state of the mean-field Hamiltonian with mean fields at time $t$.
From Eq.~\eqref{eq:app-rho_g}, $\rho_g(t)$ is written as $[\rho_g(t)]_{ll'}=\sum_{\lambda=N+1}^{2N} [U(t)]_{l\lambda}[U(t)]_{l'\lambda}^*$, where $U(t)$ is a unitary matrix to diagonalize $\mathcal{H}_{\rm MF}(t)$ with $\rho(t)$.
For the time evolution under the relaxation time approximation, we need to diagonalize $\mathcal{H}_{\rm MF}(t)$ at each time step.
The relationship between the procedure based on this approximation and that based on the Gorini-Kossakowski-Sudarshan-Lindblad (GKSL) equation~\cite{Breuer2007,manzano2020rev,Campaioli2024rev} is given in Appendix~\ref{app:GKSL}.
Note that the former considers relaxations to the ground state of an instantaneous mean-field Hamiltonian, which contrasts with previous studies incorporating relaxations to a finite-temperature equilibrium state~\cite{Reinhard2015} and a time-independent ground state for the initial state~\cite{PhysRevLett.127.127402}.
The present approach introduced in Eq.~\eqref{eq:RTA} allows us to reproduce relaxations to a metastable state with excited visons.
% The validity of this method will be discussed in Sec.~\ref{sec:discussion}.
More detailed discussions will be provided in Sec.~\ref{sec:discussion}.

\subsection{Visons and Majorana zero modes}
\label{sec:quasiparticles}

In the present calculations, we focus on spatial distributions of vison excitation and Majorana zero modes.
The expectation value $\means{W_p}$ at plaquette $p$ is evaluated for the time-dependent wave function $\kets{\Psi(t)}$.
To visualize the spatial dependence of Majorana zero modes, we introduce the local Majorana DOS at site $j$ as~\cite{Harada2023},
\begin{align}
    g_j(\varepsilon,t)=2\sum_{\lambda=1}^N |U_{j\lambda}(t)|^2\delta(\varepsilon-\varepsilon_{\lambda}).
\end{align}
Note that $\varepsilon_{\lambda}$ depends on $t$ as it is an eigenvalue of $\mathcal{H}_{\rm MF}(t)$, and $g_j(\varepsilon,t)$ satisfies the sum rule $\int_0^\infty g_j(\varepsilon,t)d\varepsilon =1$.
From the local Majorana DOS, we introduce the low-energy Majorana DOS below $\varepsilon_c$ as
\begin{align}\label{eq:low-E-Majorana-DOS}
g_j^{\rm low}=\int_0^{\varepsilon_c} g_j(\varepsilon,t) d\varepsilon. 
\end{align}
In the following calculations, we fix the cutoff to $\varepsilon_c/J=10^{-3}$.

%%%%%%%%%%%
% Result

\section{Result}
\label{sec:result}

\subsection{Effect of energy dissipation}
\label{energy-dissipation}

\begin{figure*}[t]
  \begin{center}
  \includegraphics[width=2\columnwidth,clip]{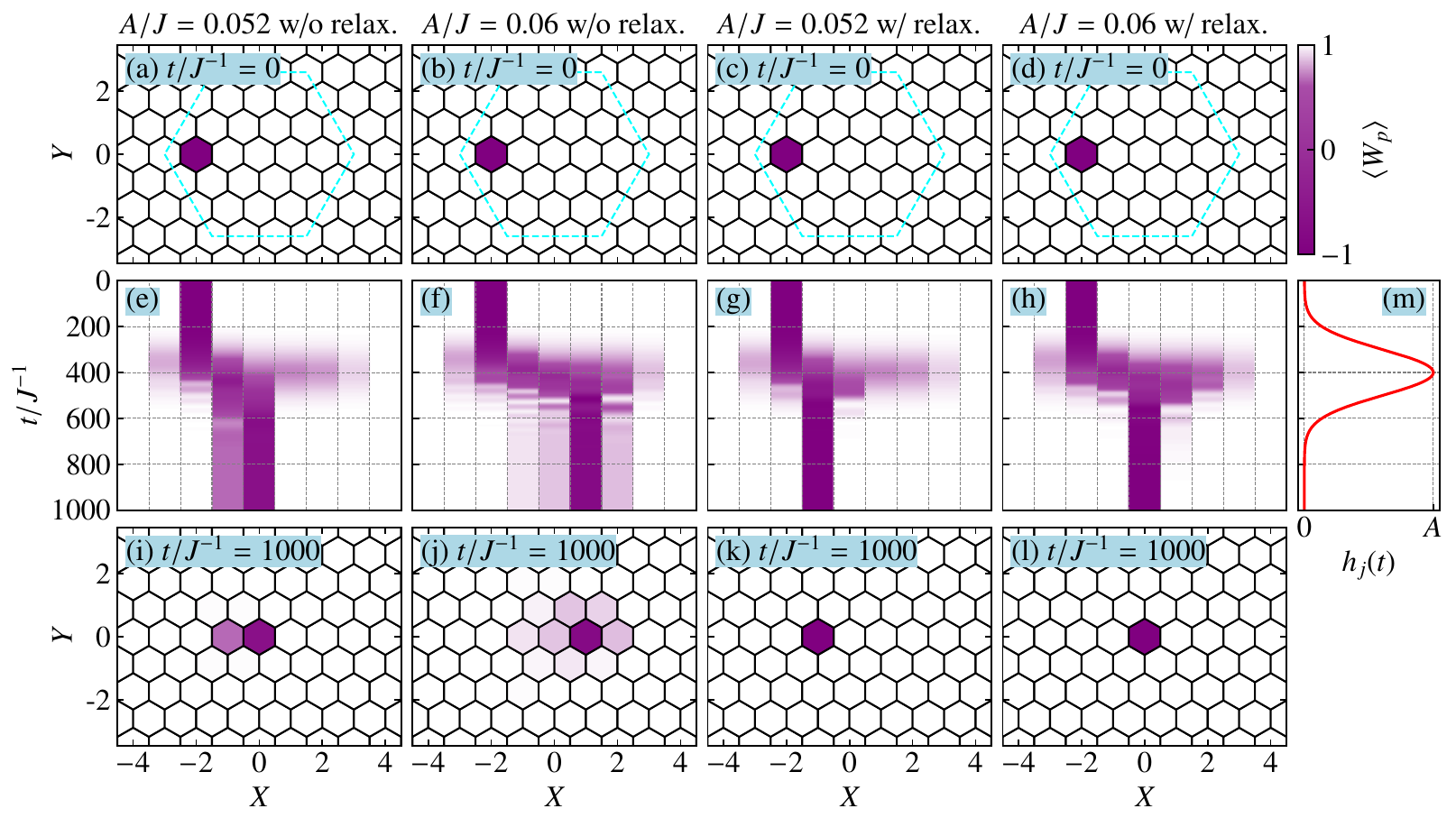}
  \caption{
  Time evolution of the spatial distributions of visons for the case where a vison is driven by time-dependent local magnetic fields, with and without the relaxation term in Eq.~\eqref{eq:vNeq-diss}, on the cluster with $N=1176$.
  [(a),(e),(i)] Color maps of the spatial distributions of visons at (a) $t/J^{-1}=0$, (i) $t/J^{-1}=1000$, and (e) the time evolution of $\means{W_p}$ along the line $Y=0$ for $A/J=0.052$ without the relaxation term.
  [(b),(f),(j)] Corresponding plots for $A/J=0.06$ without the relaxation term.
  (c)--(l) Corresponding plots for [(c),(g),(k)] $A/J=0.06$ and [(d),(h),(l)] $A/J=0.06$, calculated in the presence of the relaxation term with $\tau/J^{-1}=50$.
  (m) Time dependence of $h_j(t)$ applied to sites inside the area $\mathcal{S}$ surrounded by the dashed cyan lines presented in (a)--(d).
  The other parameters are set to $\kappa/J=0.1$, $t_c/J^{-1}=400$, and $\sigma/J^{-1}=100$.
  }
  \label{fig:Wp_move_hikaku}
  \end{center}
\end{figure*}

\begin{figure*}[t]
  \begin{center}
  \includegraphics[width=2\columnwidth,clip]{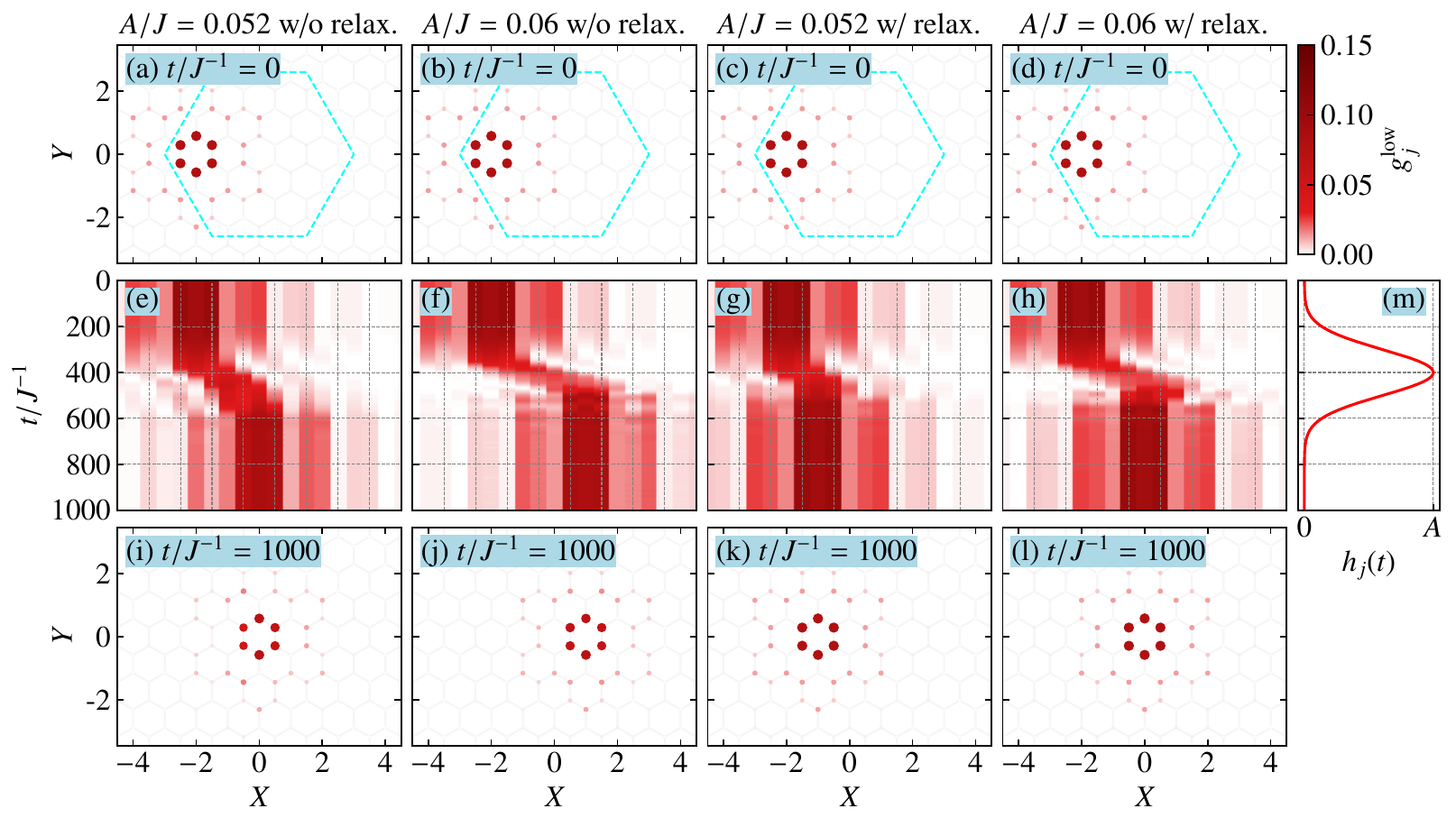}
  \caption{
  Corresponding plots to Fig.~\ref{fig:Wp_move_hikaku} for the spatial distributions of the low-energy local Majorana DOS $g_j^{\rm low}$.
  (e)--(h) show the color maps of $g_{X}(t)=\frac{1}{2}\sum_{|Y|<Y_c}g_{j=(X,Y)}^{\rm low}$ on the plane of $X$ and $t$ with $Y_c= 6/\sqrt{3}$.
  Note that the summation range corresponds to the plot range of the $Y$ axis in (a)--(d) and (i)--(l).
  }
  \label{fig:ldos_move_hikaku}
  \end{center}
\end{figure*}

In this section, we examine the effect of the relaxation term in Eq.~\eqref{eq:vNeq-diss}.
We assume that a vison with $W_p=-1$ is excited at the hexagon centered at $(X,Y)=(-2,0)$, as presented in Fig.~\ref{fig:Wp_move_hikaku}(a).
To drive the vison to the right direction, we introduce the time-dependent magnetic fields $h_j(t)$ applied to sites $j\in \mathcal{S}$, where $\mathcal{S}$ is represented by the cyan hexagon centered at $(X,Y)=(0,0)$ in Fig.~\ref{fig:Wp_move_hikaku}(a).
The time dependence of the local field is assumed to be Gaussian-type as
\begin{align}
    h_j(t)=A \exp\left[-\frac{(t-t_c)^2}{2\sigma^2}\right],
\end{align}
as shown in Fig.~\ref{fig:Wp_move_hikaku}(m).
Note that this local field is independent of site $j$ inside $\mathcal{S}$.
First, we consider the case without the relaxation term, corresponding to $\tau=\infty$ in Eq.~\eqref{eq:RTA}.
Figures~\ref{fig:Wp_move_hikaku}(a), \ref{fig:Wp_move_hikaku}(e), and \ref{fig:Wp_move_hikaku}(i) show the results for the time evolution of the spatial distributions of visons for the system with $\kappa/J=0.1$ on the cluster with $N=1176$, and the parameters of $h_j(t)$ are set to $A/J^{-1}=0.052$, $t_c/J^{-1}=400$, and $\sigma/J^{-1}=100$.
As presented in Fig.~\ref{fig:Wp_move_hikaku}(e), the vison distribution spreads out by introducing the time dependent magnetic field and maximally extended at $t\sim 400 J^{-1}$.
The distribution of vison becomes small with decreasing the value of $h_j(t)$.
In the absence of local magnetic fields, $W_p$ on each hexagonal plaquette is a conserved quantity, and hence the vison distribution is almost unchanged for $t\gtrsim 700J^{-1}$.
We find that the vison distribution appears to be extended over two plaquettes even at $t/J^{-1}=1000$, as presented in Fig.~\ref{fig:Wp_move_hikaku}(i).
Note that there is a driving force to a vison to the right side due to the asymmetry of the spatial distribution of magnetic field centered at the initial position of a vison.
This result is consistent with our previous study, where a vison moves to the center of the region to which a time-dependent magnetic field is applied~\cite{Harada2023}.

To confirm the field-driving phenomenon of visons with other parameters, we perform the same calculation for systems with $A/J^{-1}$ changing to $0.06$.
The results are shown in Figs.~\ref{fig:Wp_move_hikaku}(b), \ref{fig:Wp_move_hikaku}(f), and \ref{fig:Wp_move_hikaku}(j).
The time evolution of the vison distribution until $t/J^{-1}\sim 400$ is similar to that for $A/J^{-1}=0.052$, but the shift of the vison position after the disappearance of the local magnetic field is larger than the case with $A/J^{-1}=0.052$.
Furthermore, Fig.~\ref{fig:Wp_move_hikaku}(j) indicates that vison distribution in the final state remains spread out.
These results are considered to originate from the enhancement of the magnetic field driving the vison and suggest that the visons driven by the field are no longer localized excitations.

Next, we investigate the time evolution of vison distributions with energy dissipation, in the presence of $I(\rho)$ in Eq.~\eqref{eq:vNeq-diss}, which is introduced as the relaxation term $I^R(\rho)$ given by applying the relaxation time approximation presented in Eq.~\eqref{eq:RTA} with a finite $\tau$.
Here, we set the relaxation time to $\tau/J^{-1}=50$~\footnote{This is estimated to $\tau\sim 3.8~{\rm ps}$ for $J=100~{\rm K}$. This is comparable with the relaxation time used in Ref.~\cite{Kanega2021}, which also discusses the effect of energy dissipation in the Kitaev model.}, except for numerical simulations on pair creation of visons in Sec.~\ref{sec:pair-creation}.
Figures~\ref{fig:Wp_move_hikaku}(c), \ref{fig:Wp_move_hikaku}(g), and \ref{fig:Wp_move_hikaku}(k) show the time evolution of the vison excitation with energy dissipation for the same system as that presented in Figs.~\ref{fig:Wp_move_hikaku}(c), \ref{fig:Wp_move_hikaku}(g), and \ref{fig:Wp_move_hikaku}(k) except for the presence of the relaxation term $I^R(\rho)$.
The time evolution of the vison distribution is almost unchanged from the case without the relaxation term until the peak time $t_c$ of the local field.
However, one can observe a substantial difference between Figs.~\ref{fig:Wp_move_hikaku}(e) and \ref{fig:Wp_move_hikaku}(g) after $t_c$.
The shift of the vison position driven by the local field is smaller than that without the relaxation term, which can be attributed to the effect of energy dissipation.
Moreover, for the case with a finite relaxation time, the spatial extension of the vison distribution is strongly suppressed, and the vison is localized at one plaquette after enough time has passed from the application of the local field, as shown in Fig.~\ref{fig:Wp_move_hikaku}(k).
In the absence of a local magnetic field, the state where a vison is localized in a plaquette is an eigenstate of the Hamiltonian, and it is expected that if the number of visons is the same, the energy will be higher when the vison distribution spreads out~\cite{Lahtinen2011}.
Therefore, by introducing energy dissipation, relaxation occurs towards a state with a vison localized in a plaquette, which is the eigenstate in the absence of $\mathcal{H}_h(t)$.
On the other hand, if there is no relaxation term, the vison distribution remains spread out even as time elapses because each vison excitation is a conserved quantity.
Hence, we conclude that energy dissipation plays an important role in driving the visons while maintaining them as localized excitations.

Figures~\ref{fig:Wp_move_hikaku}(d), \ref{fig:Wp_move_hikaku}(h), and \ref{fig:Wp_move_hikaku}(l) show the time evolution of vison distribution for $A/J=0.06$ in the presence of energy dissipation.
In this case, the vison is also localized in a plaquette when enough time has passed from the application of the local field, which is in contrast to the case without the relaxation term shown in Fig.~\ref{fig:Wp_move_hikaku}(j).
Furthermore, we observe a small shift of the vison position compared with the case without the relaxation term, which is due to energy dissipation, as discussed before.

In the presence of the uniform effective field $\kappa$, a Majorana zero mode is associated with a vison excitation~\cite{Kitaev2006,Lahtinen2011,Lahtinen2012}.
Here, to clarify the behavior of a Majorana zero mode bounded by a vison driven by the time-dependent local field in the real time simulation, we calculate the time evolution of the low-energy Majorana DOS given by Eq.~\eqref{eq:low-E-Majorana-DOS}.
Figure~\ref{fig:ldos_move_hikaku} shows the color map of the low-energy Majorana DOS calculated for the parameters used in Fig.~\ref{fig:Wp_move_hikaku}.
At $t=t_{\rm in}=0$, a Majorana zero mode is localized at the vison excitation due to the presence of the effective field $\kappa$, as shown in Figs.~\ref{fig:ldos_move_hikaku}(a)--\ref{fig:ldos_move_hikaku}(d).
From these figures, we find that a Majorana zero mode appears to follow the change of the vison position in the time evolution regardless of the presence of energy dissipation.
However, by observing the detailed shape of the Majorana zero mode, we find that its shape is deformed in the absence of the relaxation term.
Figure~\ref{fig:ldos_move_hikaku}(i) shows the spatial distribution of the low-energy Majorana DOS at the final state when $A/J = 0.052$.
From this figure, it is evident that the distribution is asymmetric. 
This asymmetry originates from the spatial extension of a vison, as shown in Fig.~\ref{fig:Wp_move_hikaku}(i).

Figures~\ref{fig:ldos_move_hikaku}(c), \ref{fig:ldos_move_hikaku}(g), and \ref{fig:ldos_move_hikaku}(k) show the calculation results for $A/J = 0.052$ in the system with the relaxation term.
It can be observed that, in the final state, the low-energy Majorana DOS exhibits a spatial distribution with sixfold symmetry, almost identical to the initial state, in contrast to Fig.~\ref{fig:ldos_move_hikaku}(i).
This symmetric distribution is attributed to the localization of the vison within a single plaquette.
Similarly, at $A/J = 0.06$, the Majorana zero mode almost completely follows the motion of the vison, as observed at $A/J = 0.052$.
On the other hand, the calculation results without considering relaxation at $A/J = 0.06$ exhibit a slightly smaller Majorana DOS in the final state, as indicated in Fig.~\ref{fig:ldos_move_hikaku}(j), in comparison with the initial state.
This is due to the larger spatial extent of the vison, as shown in Fig.~\ref{fig:Wp_move_hikaku}(j).
From these results, we conclude that the introduction of energy dissipation allows the Majorana zero mode to follow the vison without deformation.

\subsection{Size dependence of local field application area}
\label{sec:size-dep}

\begin{figure*}[t]
  \begin{center}
  \includegraphics[width=2\columnwidth,clip]{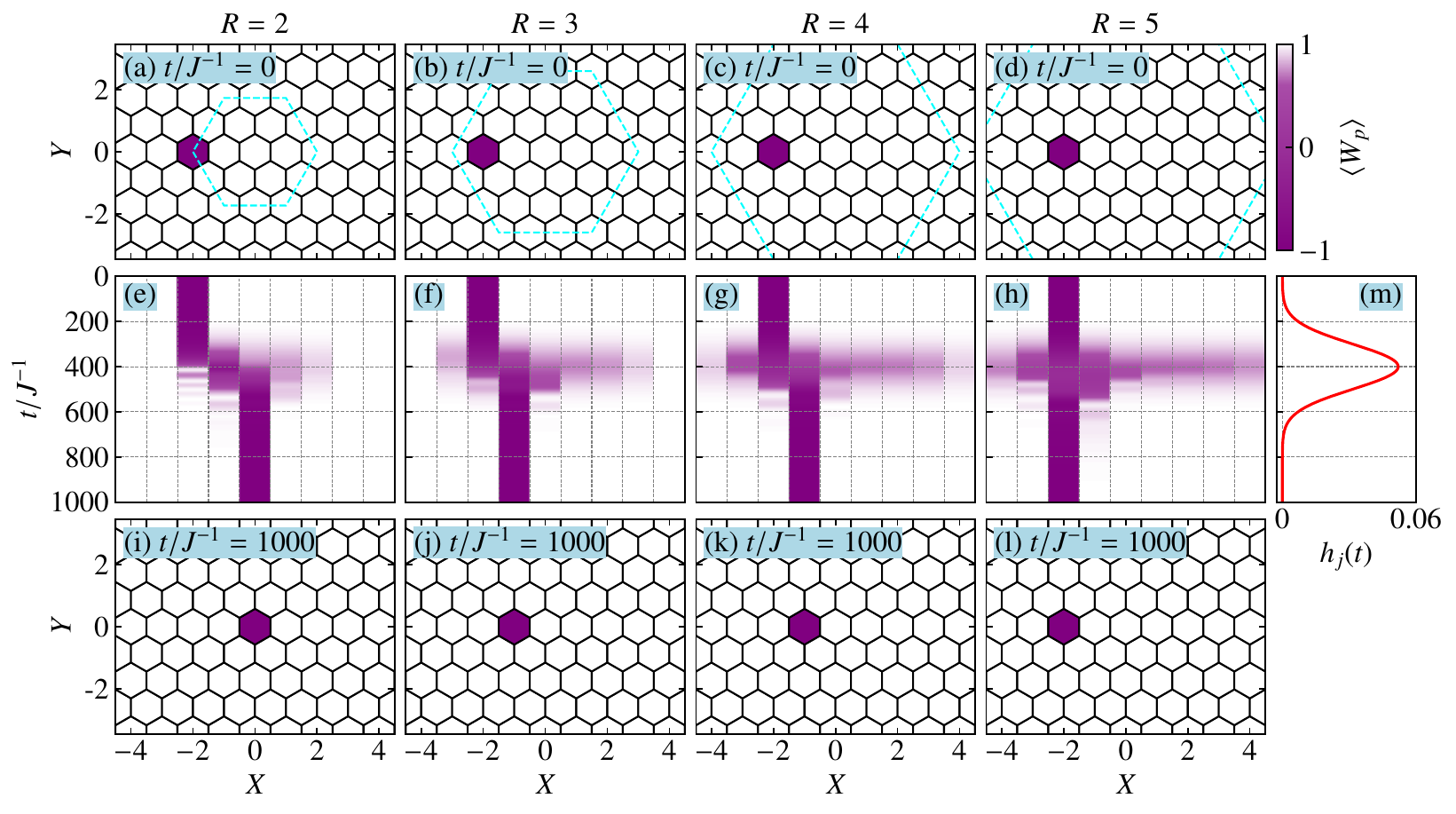}
  \caption{
  Dependence on the size $R$ of the region with a time-dependent field application in time evolution of the spatial distributions of visons, with the relaxation term on the cluster with $N=1176$, where $R$ represents the radius of the circle inscribed within the hexagon displayed by cyan in (a)--(d) to which the local field is applied.
  [(a),(e),(i)] Color maps of the spatial distributions of visons at (a) $t/J^{-1}=0$, (i) $t/J^{-1}=1000$, and (e) the time evolution of $\means{W_p}$ along the line $Y=0$ for $R=2$.
  (b)--(d),(f)--(h),(j)--(l) Corresponding plots for [(b),(f),(j)] $R=3$, [(c),(g),(k)] $R=4$, [(d),(h),(l)] $R=5$.
  (m) Time dependence of $h_j(t)$, applied to sites inside the area $\mathcal{S}$ surrounded by the dashed cyan lines presented in (a)--(d).
  The parameters used in these simulations are set to $\kappa/J=0.1$, $\tau/J^{-1}=50$, $A/J=0.052$, $t_c/J^{-1}=400$, and $\sigma/J^{-1}=100$.
  }
  \label{fig:Wp_move}
  \end{center}
\end{figure*}

\begin{figure*}[t]
  \begin{center}
  \includegraphics[width=2\columnwidth,clip]{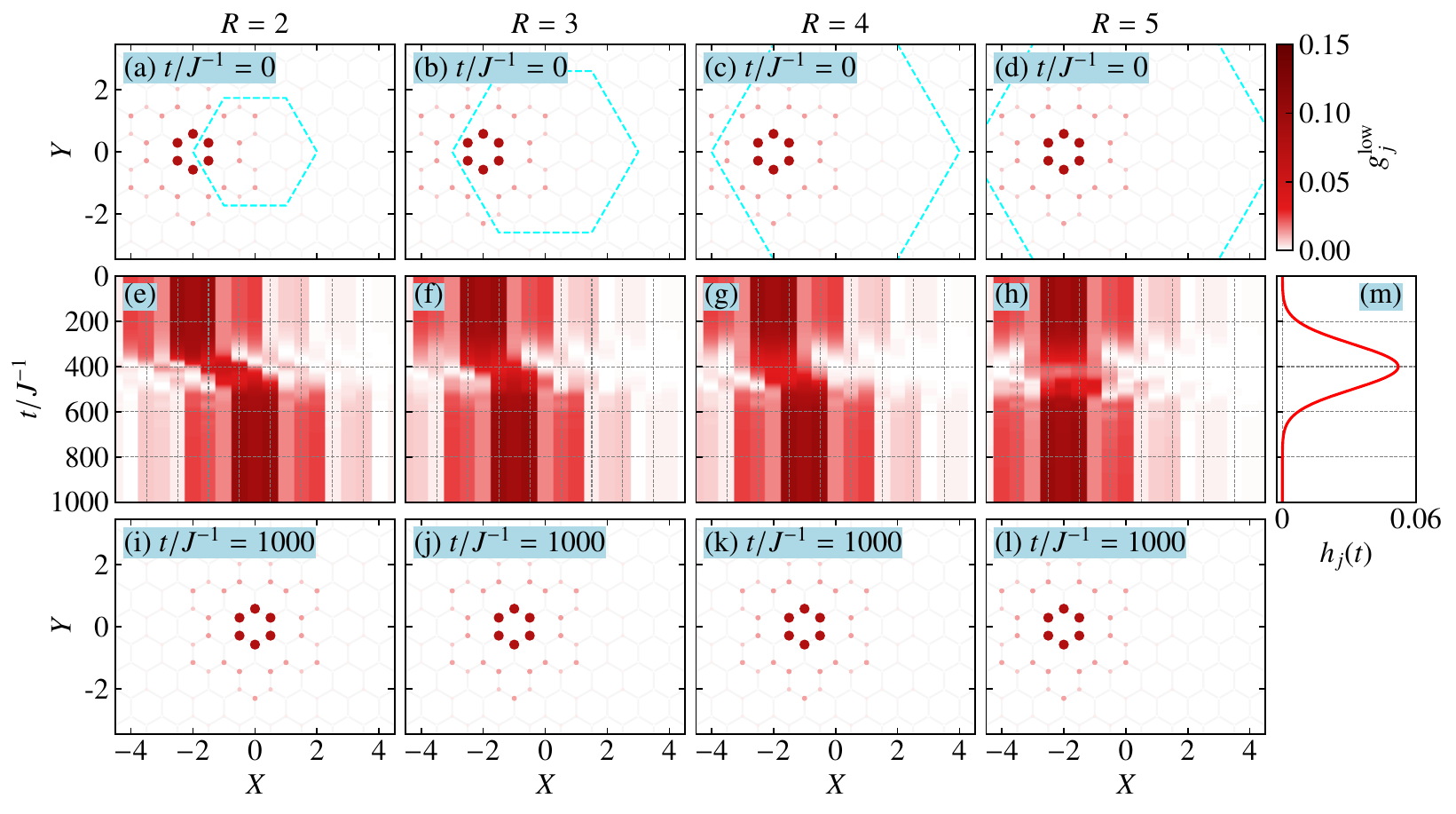}
  \caption{
    Corresponding plots to Fig.~\ref{fig:Wp_move} for the spatial distributions of the low-energy local Majorana DOS $g_j^{\rm low}$.
  (e)--(h) show the color maps of $g_{X}(t)=\frac{1}{2}\sum_{|Y|<Y_c}g_{j=(X,Y)}^{\rm low}$ on the $X$-$t$ plane, with $Y_c= 6/\sqrt{3}$.
  }
  \label{fig:ldos_move}
  \end{center}
\end{figure*}

Next, we examine the size dependence of the region $\mathcal{S}$ in Eq.~\eqref{eq:tdep-Hmag} where the time dependent magnetic field $h_j(t)$ is applied.
Figure~\ref{fig:Wp_move} presents the results for the time evolution of vison distributions.
In these calculations, we consider a single vison excitation in the hexagon centered at $(X,Y)=(-2,0)$, and the time-dependent magnetic field is applied to the sites inside a hexagon inscribed within a circle with radius $R$ centered at the origin.
Near $t_c$ with the local field taking a peak, the vison excitation spreads out significantly.
As shown in Fig.~\ref{fig:Wp_move}(e)--\ref{fig:Wp_move}(h), the spatial distribution becomes larger with increasing $R$ and extends over the region where the local magnetic field is applied.
This is attributed to the expansion of the region where $W_p$ does not commute with $H_h(t)$ and the enhancement of energy injection by the local magnetic field.
Here, we focus on the final state at $t/J^{-1}=1000$ presented in Figs.~\ref{fig:Wp_move}(i)--\ref{fig:Wp_move}(l).
The local magnetic field with $R=2$ causes the vison to shift by two plaquettes to the right, and the vison shifts by one plaquette for the cases of the $R=3$ and $4$ cases.
For the case with $R=5$, the vison remains stationary under the application of the local magnetic field.
These results suggest that the spatial modulation of the magnetic field, rather than the energy injection into the system, is crucial for driving visons.
At $R=2$, a local magnetic field is applied to two of the six sites of the plaquette where a vison is excited, while it is not applied to the other four sites.
This results in a significant asymmetry of the magnetic field applied to the vison, which can be a large driving force for the vison.
However, the importance of the spatial asymmetry of the magnetic field applied to the edge sites of the hexagon with a vison contradicts the fact that such vison driving also occurs at $R=3$ and $4$.
At $R=3$ and $4$, the same magnetic field is applied to all sites of the plaquette with the vison, and the vison itself does not detect a spatially asymmetric magnetic field.
This implies that the origin of the vison driving is not attributed to the spatial modulation of the magnetic field applied to spins composing $W_p$ with a vison.

To understand the origin of vison driving, let us focus on the spatial distribution of a Majorana zero mode.
Figure~\ref{fig:ldos_move} shows the time evolution of the local Majorana DOS for the systems corresponding to the calculations presented in Fig.~\ref{fig:Wp_move}. 
The spatial distribution of the Majorana DOS in the final state appears to be identical to a spatial shift of that in the initial state, with the shift width depending on $R$.
From the initial state shown in Figs.~\ref{fig:ldos_move}(a)--\ref{fig:ldos_move}(d), it is found that the Majorana zero mode extends beyond the six sites of $W_p$ where the vison excitation occurs.
For $R=3$ and $4$, the Majorana zero mode spans an area larger than the hexagon of $W_p$, allowing it to detect the spatial modulation of the magnetic field.
In contrast, for $R=5$, the time-dependent field application region covers almost all of the Majorana zero mode.
In this case, the spatial movement of the vison does not occur.
These results suggest that the vison driving is ascribed to the sensitivity of the Majorana zero mode to the spatial modulation of the magnetic field.

\begin{figure}[t]
  \begin{center}
  \includegraphics[width=\columnwidth,clip]{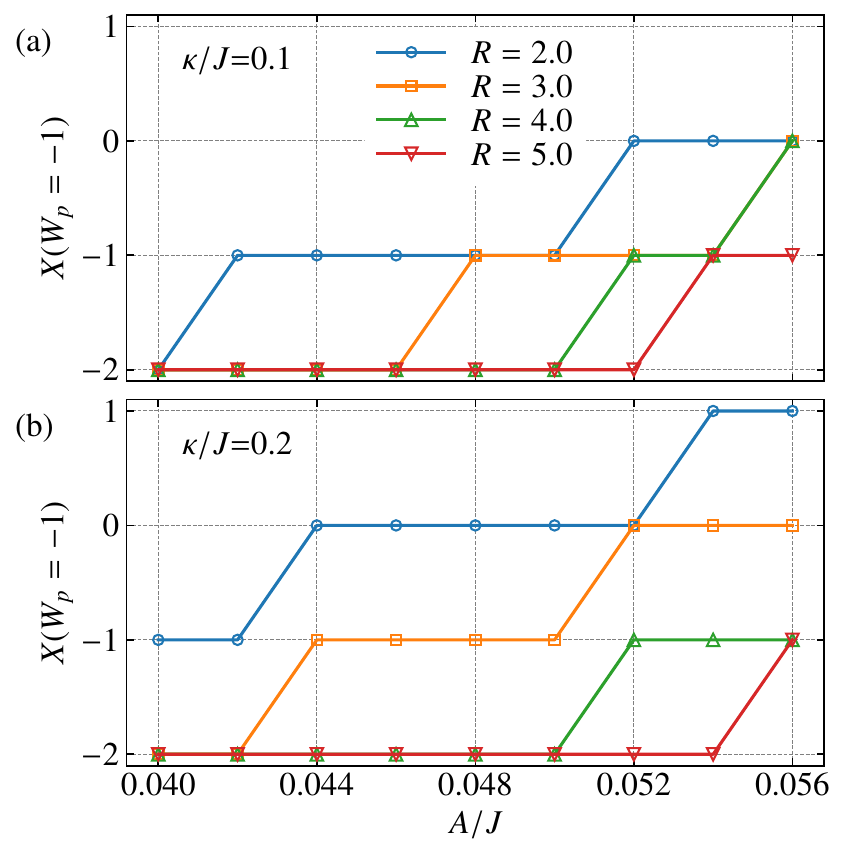}
  \caption{
  (a) $X$ coordinate of vison position $X(W_p=-1)$ in the final state at $t/J^{-1}=1000$ for systems similar to those presented in Fig.~\ref{fig:Wp_move}, as a function of the amplitude of the local field $A$.
  The result of Fig.~\ref{fig:Wp_move} corresponds to that at $A/J=0.052$.
  (b) Corresponding plot for $\kappa/J=0.2$, where the other parameters are the same as those in (a).
  }
  \label{fig:move_dep}
  \end{center}
\end{figure}

Finally, we examine the dependence of the position shift of a vison  on $A$ and $\kappa$ induced by a time-dependent local magnetic field.
Figure~\ref{fig:move_dep}(a) shows the $A$ dependence of the vison position of the final state for systems similar to those presented in Figs.~\ref{fig:Wp_move} and \ref{fig:ldos_move}.
Note that $X(W_p=-1)=-2$ corresponds to the case that vison driving by the time-dependent local field does not occur as the vison excitation is initially located in the hexagon plaquette at $(X,Y)=(-2,0)$.
As presented in Fig.~\ref{fig:move_dep}(a), the shift of the position of an excited vison becomes large with increasing $A$ for all $R$ cases.
Note that large values of $A$ lead to the occurrence of vison driving in $R=5$.
This is likely due to the small but nonzero Majorana DOS at the boundary of the magnetic field application region $\mathcal{S}$.
Let us discuss the effect of $\kappa$ on the vison-driving phenomenon.
Figure~\ref{fig:move_dep}(b) shows the result for $\kappa/J=0.2$.
Comparing this result with that in Fig.~\ref{fig:move_dep}(a) for $\kappa/J=0.1$, we find that increasing $\kappa$ facilitates vison driving for local magnetic fields with small $R$ and small $A$. 
This phenomenon is understood from the enhancement of the vison hopping amplitude with increasing $\kappa$ by a local magnetic field, which has been demonstrated in the previous studies~\cite{Joy2022,Chen2023,Harada2023}.
On the other hand, for the case with large $R$ and large $A$, the increase of $\kappa$ suppresses vison movement.
This is understood from the fact that the Majorana zero mode becomes insensitive to the spatial modulation of the magnetic field due to the reduction in the spread of the low-energy Majorana DOS with increasing $\kappa$.

\subsection{Effect of gradient field}
\label{sec:gradient-field}

Based on the calculation results presented thus far, we have found that the spread of the Majorana zero mode is highly sensitive to the spatial variation of the magnetic field.
Our findings suggest that it is possible to drive the vison even when a magnetic field is applied over a wide area, as long as the magnetic field exhibits spatial modulation.
From this consideration, we here examine effects of an gradient magnetic field on vison manipulation.
To this end, we introduce a magnetic field with spatial gradient for the $X$ direction as
\begin{align}\label{eq;gradient-field}
    h_j(t)=(A+\Delta A X_j) \exp\left[-\frac{(t-t_c)^2}{2\sigma^2}\right],
\end{align}
where $\Delta A$ is the magnitude of the gradient.
We conduct time-dependent simulations on the $N=1176$ cluster shown in Fig.~\ref{fig:lattice}(a), and the initial state is assumed to be a state with a vison excitation in the hexagon centered at $(X,Y)=(0,0)$, which differs from the previous calculations.
To avoid interference between applied magnetic fields and edge states, we apply the gradient field given in Eq.~\eqref{eq;gradient-field} to the sites inside the cyan dashed line shown in Fig.~\ref{fig:lattice}(a), which is a hexagon inscribed in a circle with $R=8$.
Namely, we choose such a site set as $\mathcal{S}$ in Eq.\eqref{eq:tdep-Hmag}.

\begin{figure*}[t]
  \begin{center}
  \includegraphics[width=2\columnwidth,clip]{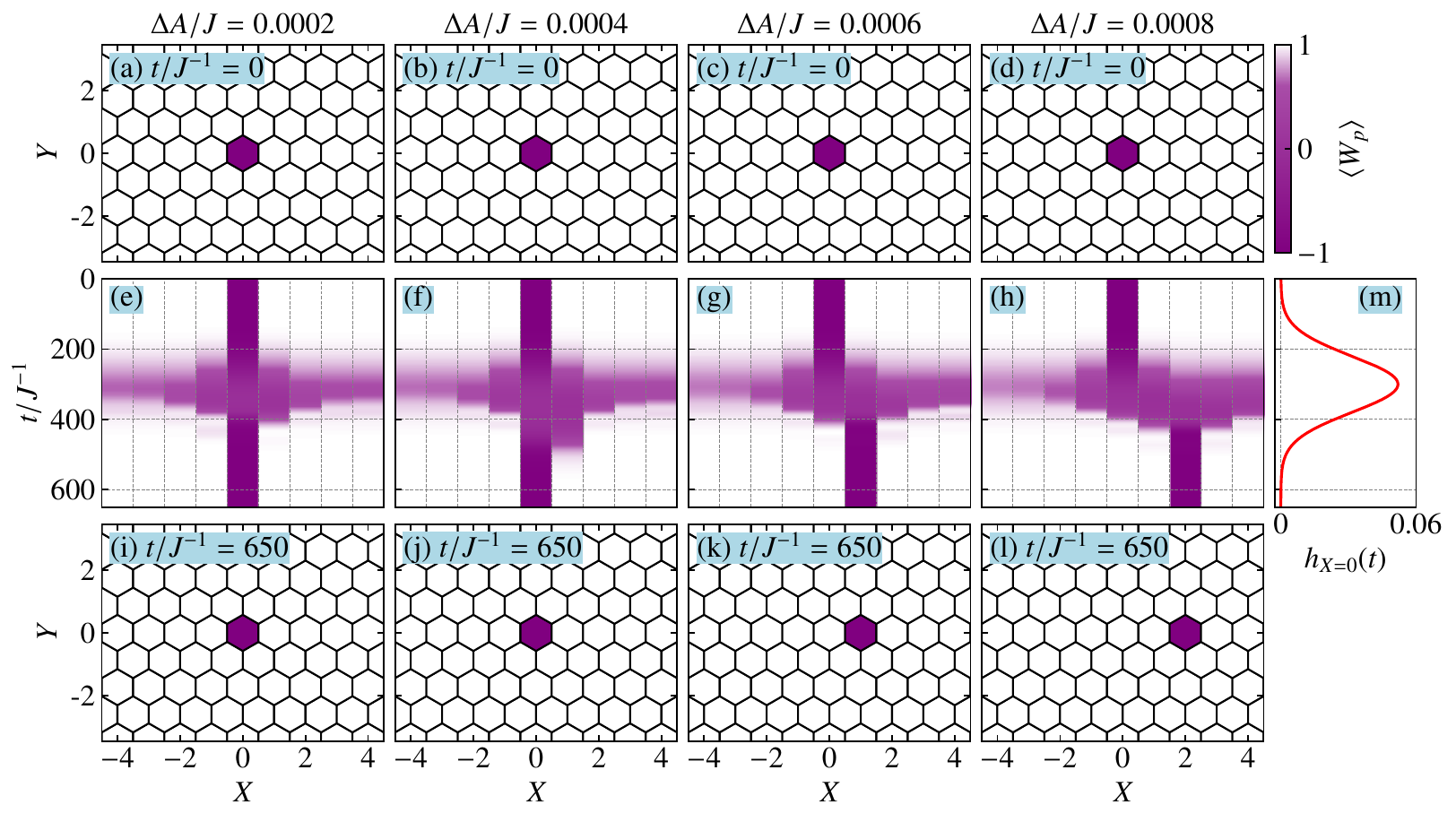}
  \caption{
  Time evolution of the spatial distributions of visons, with a relaxation term on the cluster with $N=1176$ caused by a time-dependent magnetic field with a field gradient $\Delta A$.
  [(a),(e),(i)] Color maps of the spatial distributions of visons at (a) $t/J^{-1}=0$, (i) $t/J^{-1}=1000$, and (e) the time evolution of $\means{W_p}$ along the line $Y=0$ for $\Delta A=0.0002$.
  (b)--(d),(f)--(h),(j)--(l) Corresponding plots for [(b),(f),(j)] for $\Delta A=0.0004$, [(c),(g),(k)] for $\Delta A=0.0006$, and [(d),(h),(l)] for $\Delta A=0.0008$.
  (m) Time dependence of the magnetic field applied to sites on a line with $X=0$.
  The parameters used in these simulations are set to $\kappa/J=0.1$, $\tau/J^{-1}=10$, $A/J=0.052$, $t_c/J^{-1}=300$, and $\sigma/J^{-1}=100$.
  }
  \label{fig:Wp_inc}
  \end{center}
\end{figure*}

\begin{figure*}[t]
  \begin{center}
  \includegraphics[width=2\columnwidth,clip]{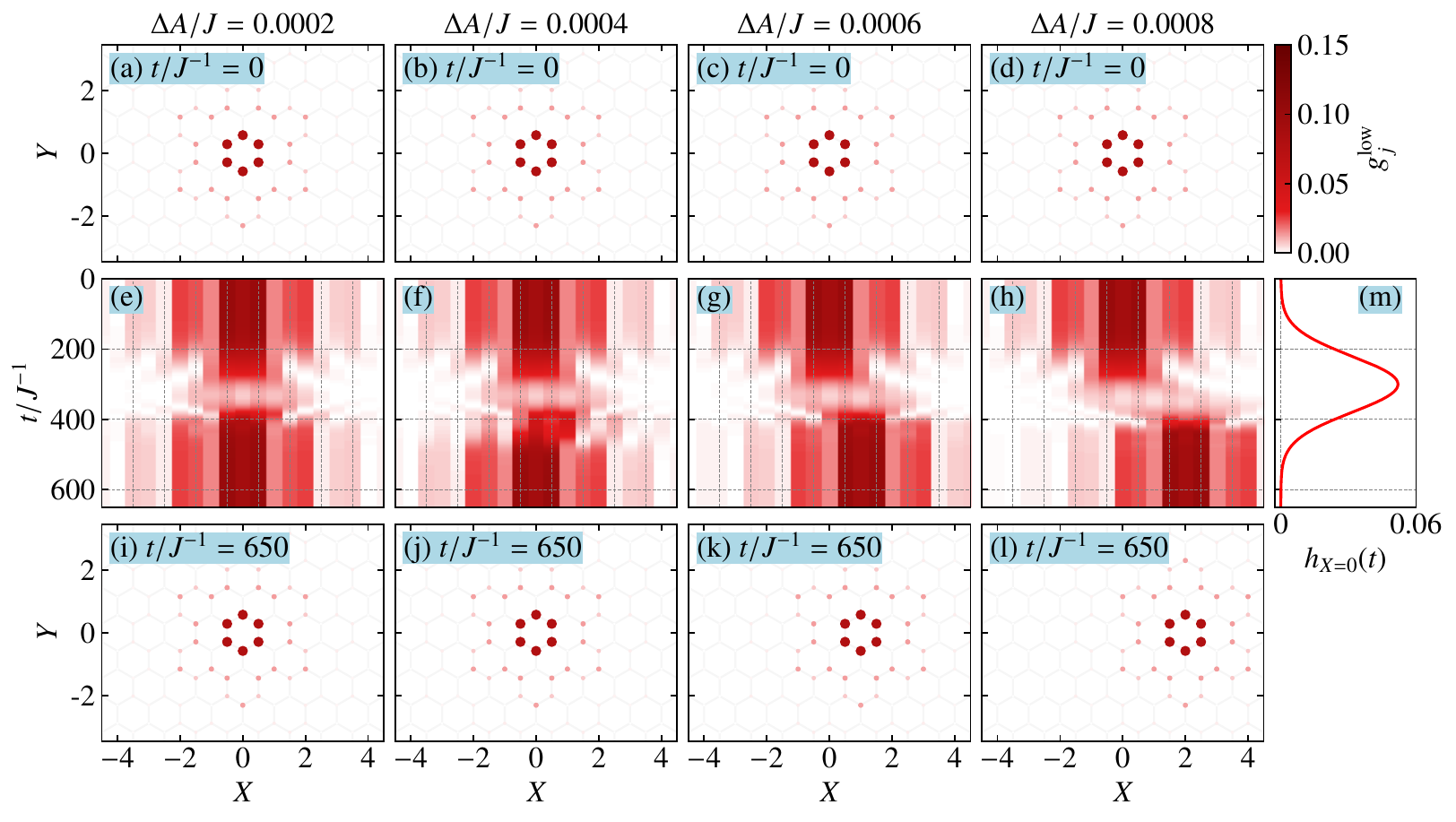}
  \caption{
    Corresponding plots to Fig.~\ref{fig:Wp_inc} for the spatial distributions of the low-energy local Majorana DOS $g_j^{\rm low}$.
  (e)--(h) present the color maps of $g_{X}(t)=\frac{1}{2}\sum_{|Y|<Y_c}g_{j=(X,Y)}^{\rm low}$ on the $X$-$t$ plane, with $Y_c= 6/\sqrt{3}$.
  }
  \label{fig:ldos_inc}
  \end{center}
\end{figure*}

Figure~\ref{fig:Wp_inc} shows the time evolution of spatial vison distributions for several values of the field gradient $\Delta A$.
As presented in Figs.~\ref{fig:Wp_inc}(e)--\ref{fig:Wp_inc}(h), the vison distribution is extended significantly around $t=t_c$ and converges to a single hexagon as time passes.
In the cases with $\Delta A=0.0002$ and $0.0004$, the position of a vison in the final state does not change from that in the initial state as shown in Figs.~\ref{fig:Wp_inc}(i) and \ref{fig:Wp_inc}(j).
For larger $\Delta A$, the vison shifts to the right side, and Figs.~\ref{fig:Wp_inc}(k) and \ref{fig:Wp_inc}(l) demonstrate that the magnitude of the shift becomes large with increasing $\Delta A$.
Furthermore, the present result indicates that the vison moves in the direction of increasing magnetic fields.
This result is understood from the asymmetry of the hopping amplitude of a vison to its left and right sides.
Previous studies have demonstrated a strong magnetic field applied to the right side of a hexagon containing a vison causes the vison to shift to the right~\cite{Joy2022,Chen2023,Harada2023}, which is consistent with the present results.

The corresponding plots to Fig.~\ref{fig:Wp_inc} for the low-energy Majorana DOS are presented in Fig.~\ref{fig:ldos_inc}.
This figure demonstrates that a vison is accompanied by a Majorana zero mode even after the application of a time-dependent magnetic field with a field gradient.
Note that the low-energy Majorana DOS appears to decrease around the peak of the time-dependent field, which may be attributed to a significant spatial distribution of vison excitations at $t\sim t_c$, as shown in Figs.~\ref{fig:Wp_inc}(e)--\ref{fig:Wp_inc}(h).

\begin{figure}[t]
  \begin{center}
  \includegraphics[width=\columnwidth,clip]{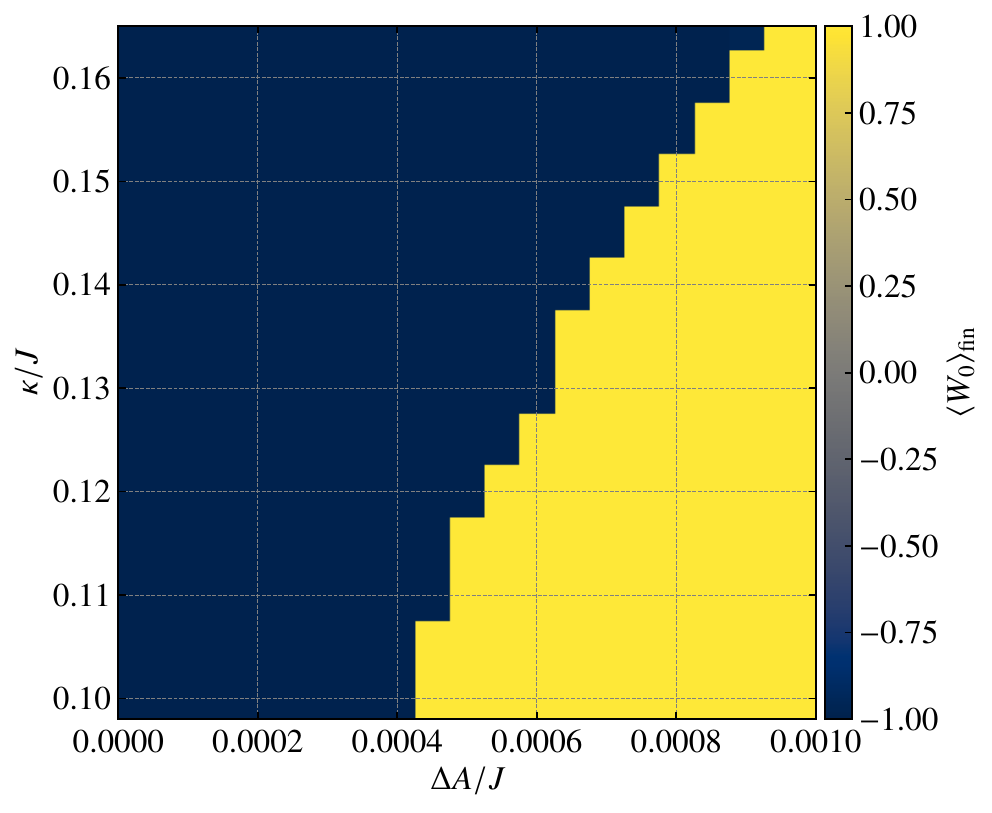}
  \caption{
    Color map of $\means{W_0}_{\rm fin}$, which is the expectation value of $W_p$ in the hexagon centered at $(X,Y)=(0,0)$ at $t/J^{-1}=650$, for the vison driving caused by a time-dependent magnetic field with a field gradient on the plane of $\Delta A$ and $\kappa$.
    The other parameters are set to the same values as those used in Figs.~\ref{fig:Wp_inc} and \ref{fig:ldos_inc}.
  }
  \label{fig:inc_cmap}
  \end{center}
\end{figure}

As shown in Figs.~\ref{fig:Wp_inc} and \ref{fig:ldos_inc}, a threshold appears to exist for $\Delta A$ in the occurrence of vison excitation driven by the gradient magnetic field.
To address this issue, we systematically examine the time evolution of the vison excitation by varying $\Delta A$ and $\kappa$.
Figure~\ref{fig:inc_cmap} displays the color map of $\means{W_0}_{\rm fin}$, defined by $\means{W_p}$ on the hexagon plaquette centered at $(X,Y)=(0,0)$ at $t/J^{-1}=650$, on the plane of $\Delta A$ and $\kappa$.
The color map is divided into two regions with $\means{W_0}_{\rm fin}=-1$ and $\means{W_0}_{\rm fin}=+1$.
The former, indicated in dark blue, corresponds to the case where the gradient field does not change the position of a vison excitation, and the latter, indicated in yellow, corresponds to the case where the gradient field drives the vison excitation.
The boundary represents the threshold for vison manipulation driven by the gradient magnetic field.
Figure~\ref{fig:inc_cmap} indicates that as $\kappa$ increases,  driving a vison excitation by a gradient magnetic field becomes more difficult.
This is due to the shrinking spatial distribution of the Majorana zero mode, resulting in the insensitivity of visons to the gradient field.
The upward trend of the threshold on the plane of $\Delta A$ and $\kappa$ suggests that magnetic fields with a smaller field gradient can drive visons when $\kappa$ is small.
This is considered important in the context of examining vison-driven mechanisms in real materials.

\subsection{Pair annihilation of visons}
\label{sec:annihilation}

\begin{figure*}[t]
  \begin{center}
  \includegraphics[width=2\columnwidth,clip]{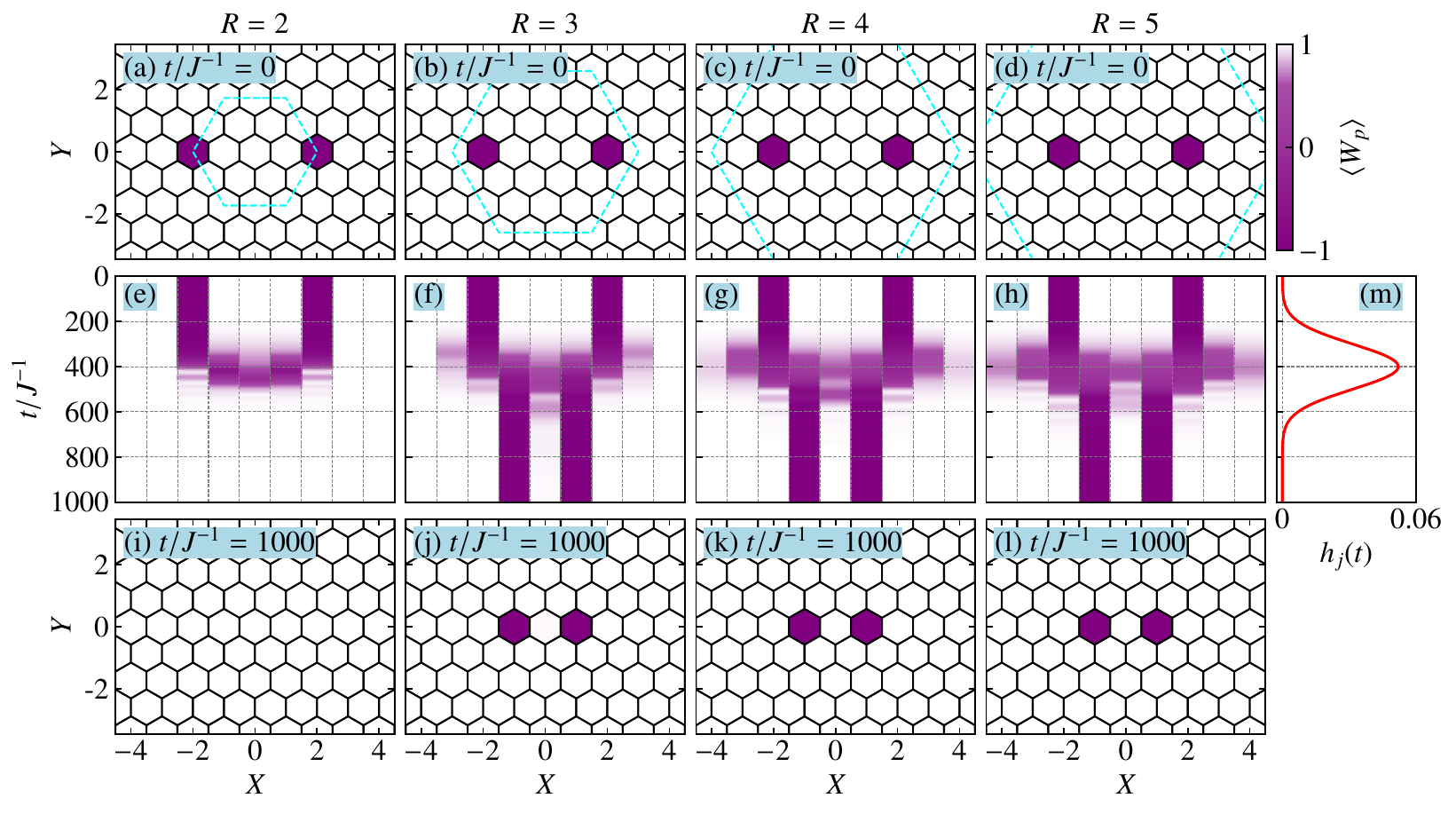}
  \caption{
  Time evolution of the spatial distributions of visons, starting from two visons excited at $(X,Y)=(\pm2,0)$, with the relaxation term on the cluster with $N=1176$, where $R$ is the radius of the circle inscribed within the hexagon displayed in cyan in (a)--(d) to which the local field is applied.
  [(a),(e),(i)] Color maps of the spatial distributions of visons at (a) $t/J^{-1}=0$, (i) $t/J^{-1}=1000$, and (e) the time evolution of $\means{W_p}$ along the line $Y=0$ for $R=2$.
  (b)--(d),(f)--(h),(j)--(l) Corresponding plots [(b),(f),(j)] for $R=3$, [(c),(g),(k)] for $R=4$, [(d),(h),(l)] for $R=5$.
  (m) Time dependence of $h_j(t)$, applied to sites inside the area $\mathcal{S}$ surrounded by the dashed cyan lines presented in (a)--(d).
  The parameters used in these simulations are set to $\kappa/J=0.1$, $\tau/J^{-1}=50$, $A/J=0.052$, $t_c/J^{-1}=400$, and $\sigma/J^{-1}=100$.
  }
  \label{fig:Wp_ann}
  \end{center}
\end{figure*}

In the previous sections, we have investigated the dynamics of visons by applying a magnetic field when a vison is present in the system.
As the next step, we explore the possibility of creating and annihilating visons by applying a magnetic field.
In this section, we examine the pair annihilation of visons in real-time simulations. 
Figure~\ref{fig:Wp_ann} shows real-time simulations with two excited visons at the initial time, driven by a time-dependent magnetic field applied to sites inside the cyan hexagons without the field gradient ($\Delta A = 0$).
In the case of $R=2$, the two visons exhibit fusion around the time when the magnetic field reaches its peak, as shown in Fig.~\ref{fig:Wp_ann}(e).
As a result, in the final state, there are no visons remaining in the system.
On the other hand, for the cases with $R=3$, $4$, and $5$, such fusion does not occur, although the two visons approach each other, driven by the local magnetic field.
These results can be explained by the reduced driving force on the visons for larger areas of the applied time-dependent magnetic field, due to the decreased spatial modulation of the magnetic field affecting the Majorana zero modes, as discussed in Sec.~\ref{sec:size-dep}.

\begin{figure}[t]
  \begin{center}
  \includegraphics[width=\columnwidth,clip]{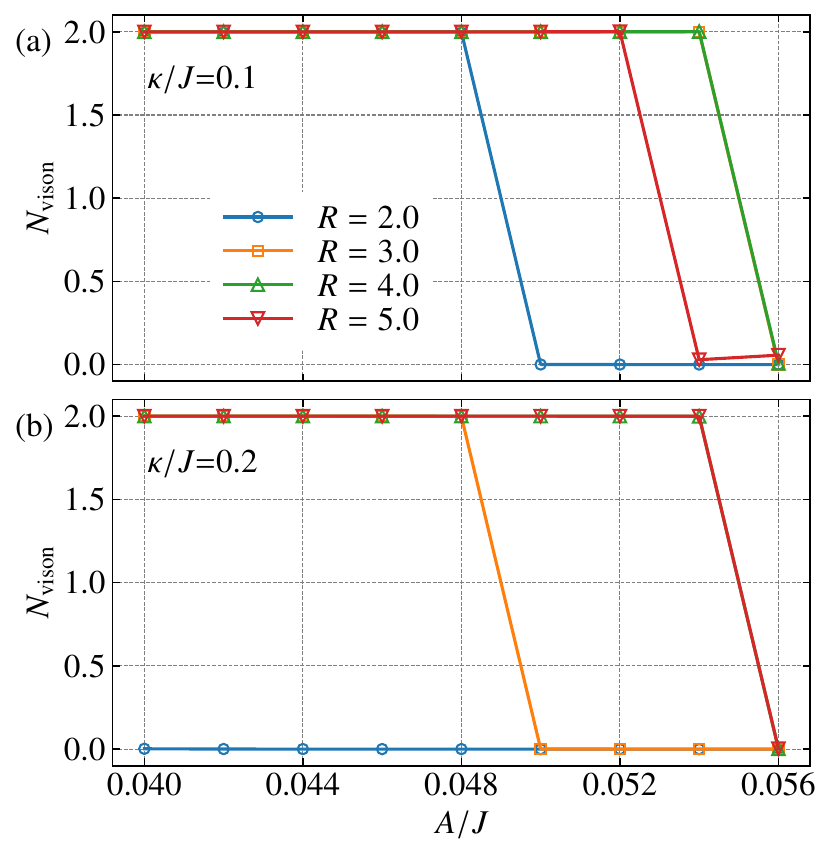}
  \caption{
  (a) Total number of visons, $N_{\rm vison}=\sum_p (1-W_p)/2$ at $t/J^{-1}=1000$ for systems similar to those presented in Fig.~\ref{fig:Wp_ann}, as a function of the amplitude of the local field $A$.
   The result of Fig.~\ref{fig:Wp_ann} corresponds to that at $A/J=0.052$.
  (b) Corresponding plot for $\kappa/J=0.2$, where the other parameters are the same as those in (a).
  }
  \label{fig:ann_dep}
  \end{center}
\end{figure}

Next, we examine how vison annihilation depends on the amplitude of time-dependent magnetic fields.
Figure~\ref{fig:ann_dep}(a) presents the $A$ dependence of the number of visons, $N_{\rm vison}$, in the final state at $t/J^{-1}=1000$, where the other parameters are set to those used in Fig.~\ref{fig:Wp_ann}.
The case with $N_{\rm vison}=2$ indicates that the number of visons has remained unchanged from the initial state, meaning that no annihilation occurs.
Conversely, when $N_{\rm vison}=0$, visons disappear from the system due to pair annihilation.
In Fig.~\ref{fig:ann_dep}(a), for $R=5$ at $A/J=0.054$ and $0.056$, $N_{\rm vison}$ is not zero despite its small value.
This is because the system has not fully relaxed even at $t/J^{-1}=1000$.
Given that we have confirmed $N_{\rm vison}$ asymptotically approaches zero over time, it is expected to become zero when considering the final state after a sufficiently long period of time.
Figure~\ref{fig:ann_dep}(a) shows that when $A$ exceeds a certain value, annihilation can occur.
This threshold increases with increasing $R$, implying that vison annihilation becomes more difficult as $R$ increases.
However, for $R=5$, the threshold is lower than that for $R=3$ and $4$.
The nonmonotonic behavior of the threshold is considered to be attributed to a significant energy injection in the case of $R=5$, which can strongly drive a vison and consequently induce annihilation.
We also present the results for $\kappa/J=0.2$ in Fig.~\ref{fig:ann_dep}(b).
In the case of $R=2$, vison pair annihilation occurs within the range depicted in this figure.
The threshold appears between $A/J=0.048$ and $0.05$ for $R=3$, and between $A/J=0.054$ and $0.056$ for $R=4$ and $5$.
These results suggest that a larger value of $\kappa$ facilitates vison annihilation, and the nonmonotonic behavior of the threshold with respect to $A/J$ observed at $\kappa/J=0.1$ does not occur at $\kappa/J=0.2$.

\subsection{Pair creation of visons}
\label{sec:pair-creation}

\begin{figure*}[t]
  \begin{center}
  \includegraphics[width=2\columnwidth,clip]{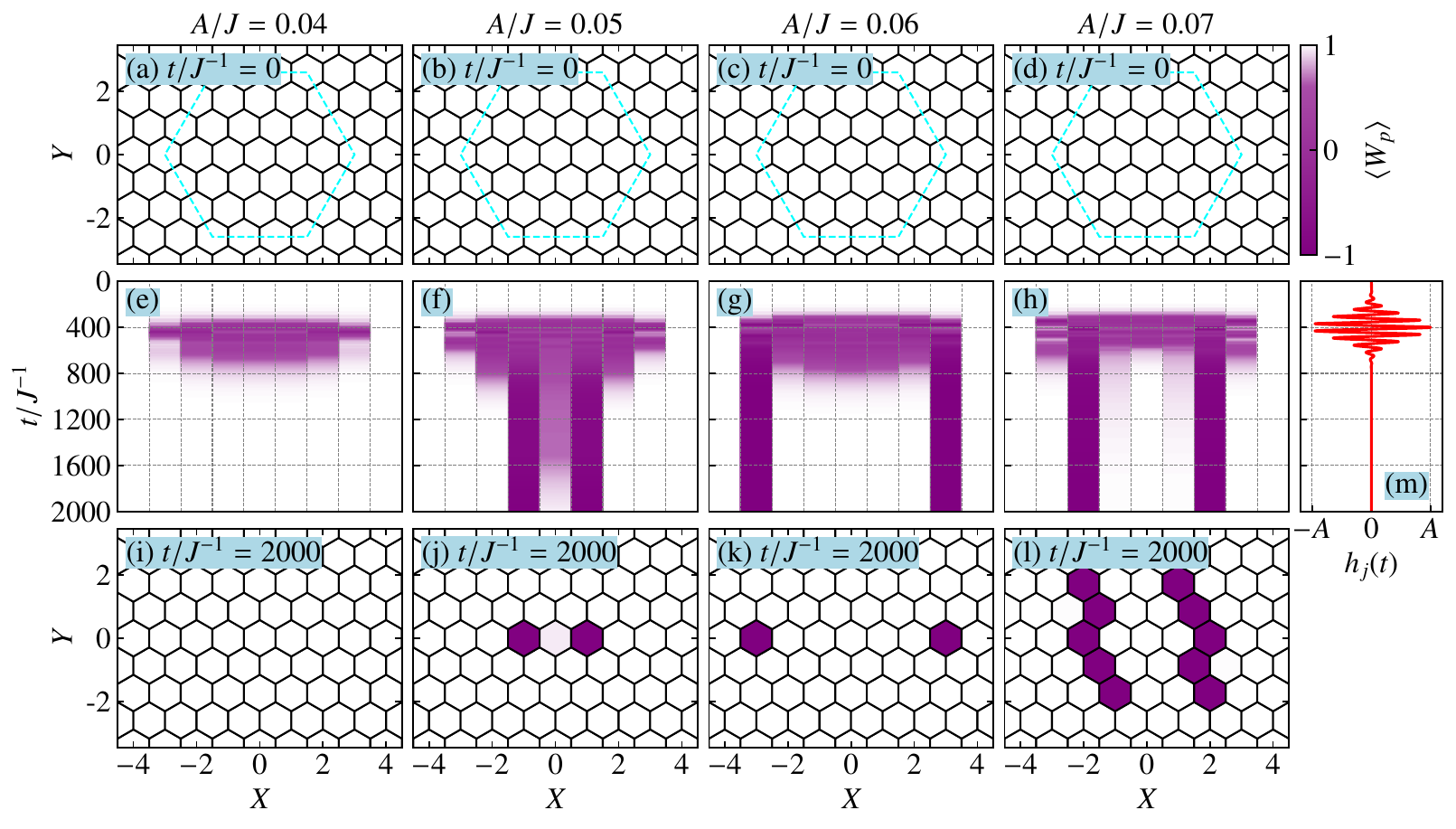}
  \caption{
    Time evolution of the spatial distributions of visons, starting from a state without visons, with the relaxation term on the cluster with $N=294$ composed of the sites inside the magenta dashed-dotted line in Fig.~\ref{fig:lattice}(a).
  [(a),(e),(i)] Color maps of the spatial distributions of visons at (a) $t/J^{-1}=0$, (i) $t/J^{-1}=2000$, and (e) the time evolution of $\means{W_p}$ along the line $Y=0$ for $A/J=0.04$.
  (b)--(d),(f)--(h),(j)--(l) Corresponding plots [(b),(f),(j)] for $A/J=0.05$, [(c),(g),(k)] for $A/J=0.06$, [(d),(h),(l)] for $A/J=0.07$.
  (m) Time dependence of $h_j(t)$, applied to sites inside the area $\mathcal{S}$ surrounded by the dashed cyan lines presented in (a)--(d).
  The parameters used in these simulations are set to $\kappa/J=0.05$, $\tau/J^{-1}=50$, $A/J=0.052$, $t_c/J^{-1}=400$, and $\sigma/J^{-1}=100$.
  }
  \label{fig:Wp_pcre}
  \end{center}
\end{figure*}

Finally, we examine the possibility of creating a vison pair triggered by time-dependent magnetic fields.
Unlike vison driving and vison pair annihilation, particular attention must be paid to  the frequency of the time-dependent magnetic field for vison creation because efficient pair creation can be achieved by resonating the frequency with the excitation energy of the visons.
It is known that the energy required to excite two visons on adjacent hexagon plaquettes is about $0.06J$ in the absence of the effective field $\kappa$~\cite{Kitaev2006,Panigrahi2023,nasu2023rev}.
Furthermore, dynamical spin correlations exhibit a coherent peak at approximately $0.1J$, originating from the two-vison excitation process~\cite{PhysRevLett.112.207203,PhysRevB.92.115127,yoshitake2016,Yoshitake2017PRBa,Yoshitake2017PRBb,Udagawa2018,Nasu2021-spin-disorder}. 
These observations suggest that visons can be resonantly excited in real-time evolution through the introduction of a time-dependent magnetic field with a frequency of approximately $0.1J$~\cite{Nasu2019_realtime,Misawa2023}.
Based on the above considerations, we introduce the following time-dependent magnetic field to attempt vison creation:
\begin{align}\label{eq:h-tdep-cos}
    h_j(t) = A \exp\left[-\frac{(t-t_c)^2}{2\sigma^2}\right]\cos\bigl(\omega(t-t_c)\bigr),
\end{align}
where we set $\omega=0.1J$.
Due to the rapid oscillation of the magnetic field, smaller time steps are necessary in numerical simulations with the above time-dependent field.
In the present calculations, we use $\Delta t=0.01J^{-1}$ unlike previous calculations.
However, this choice makes numerical computations significantly more expensive.
To reduce computational costs and complete the calculations within a realistic time scale, we set the system size to $N=294$ and the relaxation time to $\tau/J^{-1}=10$.

Figure~\ref{fig:Wp_pcre} shows the time evolution starting from a system without visons under the influence of the time-dependent field given in Eq.~\eqref{eq:h-tdep-cos}.
This field is applied to the sites surrounded by the cyan dashed lines in Fig.~\ref{fig:Wp_pcre}(a)--\ref{fig:Wp_pcre}(d), and its time dependence is displayed in Fig.~\ref{fig:Wp_pcre}(m).
Figures~\ref{fig:Wp_pcre}(a), \ref{fig:Wp_pcre}(e), and \ref{fig:Wp_pcre}(i) show the time evolution of the vison distribution for $A/J=0.04$ in Eq.~\eqref{eq:h-tdep-cos}.
As shown in these figures, although the values of $\means{W_p}$ deviate from unity around $t_c=400J^{-1}$, they return to unity in the final state at $t/J^{-1}=2000$, as shown in Fig.~\ref{fig:Wp_pcre}(e).
This result indicates that visons are not excited by the time-dependent field with $A/J=0.04$.
Figures~\ref{fig:Wp_pcre}(b), \ref{fig:Wp_pcre}(f), and \ref{fig:Wp_pcre}(j) display the results for $A/J=0.05$.
We find that, in this case, two visons are created at $(X,Y)=(\pm1,0)$ by the time-dependent field, as presented in Fig.~\ref{fig:Wp_pcre}(j).
In this figure, the value of $\means{W_p}$ at $(X,Y)=(0,0)$ does not appear to be unity.
The result indicates that the value of $\means{W_p}$ has not yet converged.
In fact, as shown in Fig.~\ref{fig:Wp_pcre}(f), the value at $X=0$ demonstrates asymptotic behavior towards 1.
Therefore, it is expected that this value will reach 1 after a sufficiently long time has passed due to the presence of energy dissipation.
Consequently, we conclude that vison pair creation occurs at $A/J=0.05$, and increasing $A$ facilitates the creation of vison pairs.

We expect that a further increase in $A$ will lead to the pair creation of visons with a long distance separation because it strengthens the driving force that moves the excited visons outward.
As presented in Fig.~\ref{fig:Wp_pcre}(k), two visons are created at $(X,Y)=(\pm3,0)$ for $A/J=0.06$.
On the other hand, in the case of $A/J=0.07$, we find that more than two visons are excited.
This result is attributed to a large injection of energy into the system.
Based on these results, it is found that selecting the appropriate intensity of the time-dependent external field is crucial for exciting a pair of visons that are spatially separated.
While applying a magnetic field over a broader area may potentially excite visons located at greater distances, this necessitates calculations with larger clusters.
Therefore, systematic investigations into this issue remain a future challenge.

%%%%%%%%%%%
% discussion

\section{Discussion}
\label{sec:discussion}

Based on the results obtained so far, we have demonstrated that visons can be driven, annihilated, and generated by magnetic fields with spatiotemporal modulations.
Additionally, we have clarified that a Majorana zero mode always accompanies a vison, even when it is driven by time-dependent magnetic fields.
Our findings suggest that it is possible to control non-Abelian anyons spatiotemporally using magnetic fields.
In this section, we discuss the feasibility of realizing the non-Abelian anyon control proposed in this study within real materials.

First, let us discuss the feasibility of vison manipulations based on the results in Sec.~\ref{sec:size-dep}, which examine the dependence of the applied area of time-dependent magnetic fields on the driving of non-Abelian anyons.
In the Kitaev candidate material $\alpha$-RuCl$_3$, the Kitaev interaction is estimated to be $J \sim 100~\rm K$~\cite{Banerjee2016_NatMate,Winter2016,Do2017majorana,hirobe2017}, and the topological gap in the Majorana fermion system has been estimated to be $\Delta_M=10~{\rm K}$ in the presence of a uniform magnetic field of $10~{\rm T}$, as determined by specific heat measurements~\cite{tanaka2022thermodynamic,Imamura2024}.
Based on these experimental results, the magnitude of the effective field in Eq.~\eqref{eq:Kitaev-Hamil} is estimated to be $\kappa/J \lesssim 0.1$ from the relation $\Delta_M=3\sqrt{3}\kappa/4$.
Figure~\ref{fig:ldos_move_hikaku} suggests that the radius of the spread of Majorana zero modes, denoted as $\xi$, is approximately 3 in units of the length of the primitive translation vectors when $\kappa/J = 0.1$.
Therefore, in the real material $\alpha$-RuCl$_3$~\cite{Johnson2015}, $\xi$ is expected to be 3 or more, depending on the magnitude of a uniform static magnetic field.
According to the results of numerical simulations, Fig.~\ref{fig:ldos_move} shows that even with $R = 4$ at $\kappa/J = 0.1$, it is possible to drive a vison.
Since $\kappa$ is considered to be smaller than $0.1J$ in $\alpha$-RuCl$_3$ and the radius of the Majorana zero mode is larger, we expect that a vison can be driven even when a time-dependent magnetic field is applied over a larger region than $R=5$.
If a magnetic field can be locally applied to a spatial region on the order of 1-10~nm, it is considered possible to drive a vison by such a magnetic field.
However, modulating a magnetic field on the scale of $A/J \sim 0.05$, i.e., on the order of $\sim 1~{\rm T}$, is considered difficult using current technology.
To realize local-field-driven vison manipulations, it might be necessary to select candidate materials with small Kitaev interactions~\cite{Yamada2017,Jang2020A2PrO3}, which would enable the driving of visons with a smaller magnetic field.

We also discuss the possibility of vison control using gradient magnetic fields.
From Fig.\ref{fig:inc_cmap}, the minimum value of $\Delta A$ that allows the movement of the vison by a gradient magnetic field is $\Delta A_{\rm min} = 4 \times 10^{-4}$ when $\kappa/J = 0.1$.
In $\alpha$-RuCl$_3$, this magnitude can be estimated to be $\sim 10~{\rm mT/nm}$.
To evaluate the feasibility of this gradient magnetic field, we refer to discussions regarding the driving of other quasiparticles using gradient magnetic fields.
Previous studies have investigated driving magnetic skyrmions with a magnetic field gradient.
It has been reported that skyrmions can be driven by a magnetic field gradient of several ${\rm mT/mm}$~\cite{Zhang_skyrmion_2018}.
This magnitude is $10^{-5}$ times smaller than that required for driving visons, suggesting that the schemes used for skyrmions cannot be applied to vison manipulation.
This difference arises because the size of skyrmions is significantly larger than that of Majorana zero modes localized on a vison.
Another potential method is based on the recently proposed STM technique that generates nanoscale magnetic fields using Dy atoms\cite{Singha2021localmag}.
The spatial gradient of the magnetic field produced by Dy atoms is estimated to be on the order of $10~{\rm mT/nm}$, making it a promising candidate for driving visons.
Alternatively, we can consider an approach that reduces the uniform effective magnetic field $\kappa$.
As shown in Fig.~\ref{fig:inc_cmap}, decreasing $\kappa$ increases the size of the Majorana zero modes, thereby lowering the required magnitude of the magnetic field gradient for vison manipulation.
In cases of smaller $\kappa$, it might be possible to drive visons without relying on the nanoscale magnetic fields produced by Dy atoms.

Similar to the vison driven by local and gradient magnetic fields, pair annihilation and pair creation of visons in our scheme require magnetic fields of approximately $1~{\rm T}$ with spatial modulations for the case of the Kitaev candidate material $\alpha$-RuCl$_3$.
In particular, our results suggest that an oscillating magnetic field with frequency $\omega/J=0.1$ corresponding to the vison gap facilitates pair vison creation.
Assuming $J=100~{\rm K}$ in $\alpha$-RuCl$_3$, this frequency is estimated to be approximately $0.1~{\rm THz}$.
Magnetic fields with this frequency can be generated by a laser pulse via the inverse Faraday effect~\cite{kimel2005ultrafast,Stanciu2007}.
However, since it is extremely difficult to locally irradiate light, it may be necessary to reduce $\kappa$ or use materials with small Kitaev interactions to achieve pair vison creation in real materials.
Theoretically, it is important to systematically investigate a broader range of parameter regions by varying the spatial distribution $R$, the frequency $\omega$, and the pulse width $\sigma$ to explore more realistic parameters for achieving pair creation and annihilation.

Alternatively, it may be necessary to propose external effects driving visons other than magnetic fields, including not only pair annihilation and pair creation of visons but also their spatial movement.
For example, candidates for such external fields may include modulating Dzyaloshinskii-Moriya interactions using AFM~\cite{Jang2021}, electric field effects through magnetoelectric effects~\cite{Bolens2018,Kocsis2022,Furuya2024}, thermal or strain gradients~\cite{Nasu2017,Joy2022,Chen2023,Rachel2016,hong2020extreme,du2023strain}, and potentially plasmonics~\cite{gramotnev2014nanofocusing,meinzer2014plasmonic,Koenderink2015,wei2021plasmon}.
For instance, considering vison driving by a temperature gradient, it is expected that the direction of movement will depend on the direction of the applied magnetic field.
When a magnetic field is applied in the $S^z$ direction, visons move in the $X$ direction, and thus, visons will contribute to enhancing the thermal conductivity in the $X$ direction~\cite{Joy2022,Chen2023}, which could be observed experimentally.

In discussing the potential for vison manipulation in real materials, it is necessary to consider effects other than the Kitaev interactions, such as Heisenberg and $\Gamma$ interactions~\cite{PhysRevLett.105.027204,PhysRevLett.110.097204,PhysRevLett.113.107201,PhysRevLett.112.077204,Winter2016}.
When an effective field $\kappa$ is introduced, a topological gap opens in the Majorana fermion band, suggesting that itinerant Majorana fermion systems may be relatively robust against weak additional interactions.
However, in the presence of these interactions, the vison is no longer a local conserved quantity.
Therefore, these interactions must be weaker compared to the Kitaev interaction for a vison to remain well-defined as a quasiparticle.
Moreover, since the lifetime of a vison is determined by the inverse of the strength of additional interactions, the control speed of a vison must be faster than this timescale.
To perform topological quantum computation in the presence of additional interactions, it would be necessary to employ a proper scheme that maintains the quantum state of a vison and extends its lifetime.

Next, we discuss future research directions based on our findings.
Our study relies on the Jordan-Wigner transformation, which allows us to apply a magnetic field only to the $S^z$ component, thus enabling the vison to be driven solely along the $X$ axis in real space.
To perform the braiding of visons necessary for topological quantum computation, it is crucial to compute the time evolution in a setup that allows vison movement in the $Y$ direction as well.
To this end, instead of the Jordan-Wigner transformation, we should use the method involving four Majorana fermions, which is the original approach proposed by Kitaev~\cite{Kitaev2006}.
Real-time calculations using the latter method have been studied in the context of the Kitaev quantum spin liquid under uniform magnetic fields~\cite{Cookmeyer2023}.
By exploiting this approach, it might be possible to simulate the braiding of non-Abelian anyons in real-time simulations.
During this braiding process, it is also essential to focus on the time evolution of the phase of Majorana zero modes associated with visons, which remains a challenge for future work.

Finally, we comment on the introduction of energy dissipation to the Kitaev model.
In this study, we introduce a term that facilitates relaxation to the ground state of the instantaneous mean-field Hamiltonian at each time step in Eq.~\eqref{eq:vNeq} to handle excited states with visons.
The microscopic origin of energy dissipation can be attributed, for instance, to the coupling between the spin degrees of freedom of the present system and lattice vibrations as an environment.
Given that rigorously addressing such microscopic mechanisms gives rise to computational difficulties, we employ the relaxation time approximation to phenomenologically incorporate the effects of energy dissipation.
This approach assumes that the present results do not essentially depend on the details of the microscopic relaxation processes.
While the value of the relaxation time used here is determined by referring to the previous study~\footnotemark[1], we have confirmed that the value of the relaxation time does not significantly impact the present results regarding the control of vison excitations.
Furthermore, the correspondence with the GKSL equation is presented in Appendix~\ref{app:GKSL}.
As discussed in this appendix, addressing energy dissipation based on the GKSL equation may violate the particle-hole symmetry intrinsic to Majorana fermions in the Kitaev model, which is why we employ the relaxation time approximation.
Overcoming this challenge remains a topic for future research.
While previous studies have introduced terms that enforce relaxation to the true ground state~\cite{PhysRevLett.127.127402}, our approach is fundamentally different as it can also handle relaxation to metastable states.
Therefore, the method addressing energy dissipation used in this study holds potential for application to other spin systems and electron systems.
Notably, since it does not enforce relaxation to the true ground state, it may be effective for discussing metastable states in the real-time domain in correlated electron systems~\cite{Yonemitsu2003,Seo2018,Seo2024}, whereas relaxations to a metastable state have been discussed in electron systems coupled with classical spins under Gilbert damping~\cite{Koshibae2009,ono2021ultrafast, Ono2023}.

%%%%%%%%%%%
% Summary

\section{Summary}
\label{sec:summary}

We have investigated the real-time dynamics of visons in the Kitaev model under a time-dependent magnetic field to explore the possibility of manipulating, creating, and annihilating non-Abelian anyons via external fields.
Real-time simulations have been conducted in the Kitaev model using a time-dependent mean-field theory based on the von Neumann equation.
Energy dissipation has been incorporated into these simulations as an additional term, allowing the system to relax to an instantaneous mean-field ground state.
We have discovered that a vison can be moved by a locally applied time-dependent field while retaining a Majorana zero mode, and energy dissipation localizes the vison after movement, promoting the stabilization of the vison state.
Conversely, without considering energy dissipation, the introduction of a local magnetic field causes the vison to smear due to the energy injected into the system.
We have also revealed that spatial modulation of the local magnetic field is crucial for driving a vison.
Specifically, this spatial modulation is not felt directly by the vison itself, but by the Majorana zero modes, thereby driving the vison.
Reflecting this nature, not only local magnetic fields but also gradient magnetic fields can drive visons.
We have found that when the effective magnetic field is small and the spread of the Majorana zero modes is large, the vison can be moved with a smaller field gradient.
Furthermore, we have demonstrated that using a temporally and spatially varying magnetic field, pair creation and annihilation can be achieved as real-time dynamics in addition to vison driving.
Our findings suggest that the non-Abelian anyons in the Kitaev quantum spin liquid can be created, annihilated, and driven by external fields at will.
This study not only elucidates the real-time dynamics of quasiparticle excitations in quantum spin liquids but also stimulates further research on potential applications to quantum information technology in quantum spin liquids, as these processes could serve as operational elements in topological quantum computing.

\begin{acknowledgments}
The authors thank Y.~Motome, M.~Sato, T.~Okubo, Y.~Kato, T.~Misawa, K.~Fukui, and K.~Ido for fruitful discussions.
Parts of the numerical calculations were performed in the supercomputing systems in ISSP, the University of Tokyo.
This work was supported by Grant-in-Aid for Scientific Research from
JSPS, KAKENHI Grant Nos.~JP19K03742, JP20H00122, JP20K14394, JP23K13052, and JP23K25805.
It is also supported by JST PRESTO Grant No.~JPMJPR19L5. 
\end{acknowledgments}

\appendix

\section{Relation to GKSL equation}
\label{app:GKSL}

In this appendix, we discuss the relationship between the present method using the relaxation time approximation and the approach based on the GKSL equation~\cite{Breuer2007,manzano2020rev,Campaioli2024rev}. 
In the relaxation approximation given in Eq.~\eqref{eq:RTA}, the dissipation term is added such that the system is relaxed to the ground state $\kets{\Phi_{\rm MF}(t)}$ of the instantaneous mean-field Hamiltonian ${\cal H}_{\rm MF}(t)$ under the set of the MFs, $\rho(t)$, at time $t$.
Here, we introduce the density matrix $\tilde{\rho}(t)$ on the basis that diagonalizes ${\cal H}_{\rm MF}(t)$ as,
\begin{align}
  [\tilde{\rho}(t)]_{\lambda\lambda'} =\bras{\Psi(t)}f_{\lambda'}^\dagger f_{\lambda}\kets{\Psi(t)}
  =[U(t)^\dagger\rho(t)U(t)]_{\lambda\lambda'}.
\end{align}
In this representation, $\rho_g(t)$ in Eq.~\eqref{eq:RTA} is written by
\begin{align}\label{eq:app-rho_g}
    [\tilde{\rho}_g(t)]_{\lambda\lambda'}&=\bras{\Phi_{\rm MF}(t)}f_{\lambda'}^\dagger f_{\lambda}\kets{\Phi_{\rm MF}(t)}\notag\\
    &=
    \begin{cases}
        \delta_{\lambda\lambda'} & \lambda,\lambda'=N+1,N+2,\cdots,2N\\
        0 & {\rm elsewhere}
    \end{cases}.
\end{align}
On the other hand, in the GKSL equation, the energy dissipation term is given by
\begin{align}
  I^{L}(\rho)=\frac{1}{\tau}
\sum_{\Lambda=1}^N\left[L_{\Lambda}\rho L_{\Lambda}^\dagger
-\frac{1}{2}\{L_{\Lambda}^\dagger L_{\Lambda},\rho\}
\right],
\end{align}
where we assume the jump operator to be
\begin{align}\label{eq:jump-op}  
[\tilde{L}_{\Lambda}]_{\lambda\lambda'}=\delta_{\lambda,\Lambda+N}\delta_{\lambda',\Lambda}
\end{align}
for $\Lambda=1,2,\cdots,N$. 
This jump operator is introduced in a similar manner to Ref.~\cite{Kanega2021} as it is rewritten as $[L_{\Lambda}]_{ll'}=[U\tilde{L}_{\Lambda}U^\dagger]_{ll'} = U_{l,\Lambda+N}U_{l'\Lambda}^*$ on the original basis.
This operator changes a single-particle state with positive energy $\varepsilon_\Lambda$ to a state with negative energy $-\varepsilon_\Lambda$, which constitutes the ground state.

To discuss the difference between $I^{R}(\rho)$ and $I^{L}(\rho)$, we introduce the following representations on the bases diagonalizing ${\cal H}_{\rm MF}(t)$.
\begin{align}
  \tilde{I}^{R}(\rho)=U(t)^\dagger I^{R}(\rho)U(t),\quad\tilde{I}^{L}(\rho)=U(t)^\dagger I^{L}(\rho)U(t).
\end{align}
We find that the difference can be calculated as
\begin{align}
  G_{\lambda\lambda'}&=\tau\left[\tilde{I}^{L}(\rho)-\tilde{I}^{R}(\rho)\right]_{\lambda\lambda'}\notag\\
  &=
\begin{cases}
0 & \lambda= \lambda'\\
\displaystyle
\frac{1}{2}\sum_{\Lambda=N+1}^{2N}
(\delta_{\lambda,\Lambda}\rho_{\Lambda,\lambda'}+\rho_{\lambda,\Lambda}\delta_{\lambda',\Lambda})& \lambda\ne \lambda'
\end{cases}.
\label{Lind_com3}
\end{align}
From the above results, the diagonal parts of $\tilde{I}^{L}(\rho)$ are identical to those of $\tilde{I}^{R}(\rho)$, while the off-diagonal parts are not, which is interpreted as an extension of a well-studied two-level system~\cite{Breuer2007,manzano2020rev} to a correlated Majorana system.
This finding suggests that the relaxation phenomena, dominated by the diagonal terms, capture the same effect in both cases.
On the other hand, the off-diagonal parts of $\tilde{I}^{L}(\rho)$ do not coincide with those of $\tilde{I}^{R}(\rho)$.
Note that these off-diagonal terms differ between the positive energy components with $\lambda=1, 2, \cdots, N$ and the negative energy components with $\lambda=N+1, N+2, \cdots, 2N$.
This observation suggests that the particle-hole symmetry inherent in the density matrix, expressed as $\rho_{\lambda',\lambda}=-\rho_{\lambda+N,\lambda'+N}$ for $\lambda \ne \lambda'$, may be broken by the introduction of the energy dissipation term with the jump operator given by Eq.~\eqref{eq:jump-op}. 
Therefore, in this study, we choose to use $\tilde{I}^{R}(\rho)$ for real-time simulations as it has zero off-diagonal terms and consistently satisfies the particle-hole symmetry with $\rho_{\lambda,\lambda}+\rho_{\lambda+N,\lambda+N}=1$.

\bibliography{refs}

\end{document}